\documentclass[aps,prd,amsmath,amssymb,notitlepage,10pt,superscriptaddress,preprintnumbers,nofootinbib]{revtex4-1}
\usepackage{xcolor}
\usepackage{graphicx}% Include figure files
\usepackage{dcolumn}% Align table columns on decimal point
\usepackage{bm}% bold math
\def\Mpl{M_{\rm P}}

\begin{document}

\preprint{YITP-18-25, IPMU18-0060}

\title{Integrated Sachs-Wolfe-galaxy cross-correlation bounds on the two branches of the minimal theory of massive gravity}% Force line breaks with \\
 
\author{Nadia Bolis}%
\email{bolis@fzu.cz, nbolis@ucdavis.edu}
\affiliation{Central European Institute for Cosmology and Fundamental Physics (CEICO), \\Fyzik\'{a}ln\'{i} \'{u}stav Akademie v\v{e}d \v{C}R, Na Slovance 1999/2, CZ-182 21 Praha 8, Czech Republic
}
\author{Antonio De Felice}
\email{antonio.defelice@yukawa.kyoto-u.ac.jp}
\affiliation{Center for Gravitational Physics, Yukawa Institute for Theoretical Physics, Kyoto University, 606-8502, Kyoto, Japan}
\author{Shinji Mukohyama}
\email{shinji.mukohyama@yukawa.kyoto-u.ac.jp}
\affiliation{Center for Gravitational Physics, Yukawa Institute for Theoretical Physics, Kyoto University, 606-8502, Kyoto, Japan}
\affiliation{Kavli Institute for the Physics and Mathematics of the Universe (WPI),
The University of Tokyo, 277-8583, Chiba, Japan}

\date{\today}

\begin{abstract}
The minimal theory of massive gravity (MTMG) has two branches of stable cosmological solutions: a self-accelerating branch, which, except for the mass of tensor modes has exactly the same behavior of linear perturbations as $\Lambda$CDM in general relativity (GR), and a normal branch with nontrivial behavior. We explore the influence of the integrated Sachs-Wolfe-galaxy correlation constraints on the normal branch of MTMG, which, in its simplest implementation, has one free parameter more than $\Lambda$CDM in GR (or the self-accelerating branch of MTMG): $\theta$.  This parameter is related to the graviton mass and only affects the behavior of the cosmological linear perturbation dynamics. Using 2d-mass and SDSS data, we check which values of $\theta$ lead to a positive or negative cross-correlation. We find that positive cross-correlation is achieved for a large parameter-space interval.  Within this allowed region of parameter space, we perform a $\chi^2$ analysis in terms of the parameter $\theta$, while keeping the other background parameters fixed to the best-fit values of Planck. We then infer that the normal branch of MTMG fits the data well in a nontrivial portion of the parameter space, and future experiments should be able to distinguish such a model from $\Lambda$CDM in GR (or the self-accelerating branch of MTMG).
\end{abstract}

\maketitle

%\tableofcontents

\section{\label{sec:level1}Introduction}

Thanks to cosmic microwave background (CMB) experiments, such as WMAP \cite{Hinshaw:2012aka, Bennett:2012zja} and Planck \cite{Ade:2015xua}, as well as baryon acoustic oscillation (BAO) \cite{Aubourg:2014yra,Cuesta:2015mqa,Wang:2016wjr} and large scale structure (LSS) experiments \cite{Paris:2017xme}, measurements of many cosmological parameters have become extremely accurate. However, admits this great success,  there seem to be some discrepancies in the measurement of some parameters and phenomena between high-redshift (early time) CMB data and low-redshift (late time) growth of structure data. This trend can be seen in several measurements.  The $\Lambda$CDM model implemented with general relativity (GR) has been found to be a good fit to the most recent CMB data from the Planck Collaboration. However, once the cosmological parameters are fixed by the CMB data, the evolution with GR predicts a growth structure that seems to be higher than values found in current redshift space distorsion (RSD) measurements (see e.g.\ \cite{Blake:2011rj}).  Moreover, the measurement of the Hubble constant $H_0$ and the sound horizon redshift drag $r_s$ from Planck are in tension with their late time counterparts from BAO experiments and Type Ia supernova observations \cite{Bernal:2016gxb}. In particular, because the constraint of $H_0$ and $r_s$ from CMB experiments is indirect, it is model dependent. The $H_0$ constraints are affected by both early and late time physics and on assumptions about the expansion history of the universe, such as dark energy models or alternatives, while  $r_s$ is only affected by early time physics such as the density and equation of state parameters.  On the other hand, unlike CMB experiments, combining BAO and SNeIa data can produce direct model independent measurements of $H_0$ and $r_s$ by adopting measurements which are related to the very last instants of the evolution of our universe \cite{Bernal:2016gxb,Cuesta:2014asa}.  Comparing these two sets of measurements results in a $3\sigma$ tension in the measurement of $H_0$. This tension between the direct model independent measurement and the indirect model dependent measurement may arise either from some unknown systematic errors or may indicate some deviation from the $\Lambda$CDM model with GR which has been assumed. 

The CMB measurement of $H_0$, as well as the prediction of growth of structure, which are in tension with low-redshift data, both assume GR governs the evolution of the universe. One possible resolution for this tension may therefore, be provided by considering the effects of modified gravity theories. As discussed in \cite{DeFelice:2016ufg} substituting GR with one of the two branches of the minimal theory of massive gravity (MTMG) \cite{DeFelice:2015hla,DeFelice:2015moy}, called a normal branch, seems to reduce the tension between early time CMB and late time RSD data sets, while the other branch of the same theory gives the same prediction as $\Lambda$CDM in GR. 

Adding a mass to the spin-2 graviton is perhaps one of the simplest extensions to GR that can be explored. In 2010 a nonlinear theory of massive gravity, called dRGT \cite{deRham:2010ik, deRham:2010kj} was discovered. This theory has five degrees of freedom: one scalar, two vector, and two tensor modes. Unfortunately however, this theory happens to be unstable around homogeneous and isotropic (Friedmann-Lemaitre-Robertson-Walker) cosmological backgrounds \cite{DeFelice:2012mx}. This instability renders dRGT (de Rham-Gabadadze-Tolley) incapable of producing stable cosmological solutions.  Several approaches were taken to resolve this issue. Either the homogeneity/isotropy of cosmological models had to be abandoned \cite{DAmico:2011eto,Gumrukcuoglu:2012aa}, or extra fields, which interact with gravity, had to be added to stabilize the theory \cite{DAmico:2012hia,Huang:2012pe}. Another option is to directly eliminate the unstable degrees of freedom as done in MTMG \cite{DeFelice:2015hla, DeFelice:2015moy}. MTMG explicitly breaks Lorentz invariance at cosmological scales, and by doing so puts constraints on the physical phase space of the theory which allows for only the two tensor degrees of freedom to propagate, as in GR. Since the scalar and vector modes are absent in this theory, its background evolution matches that of GR and is stable.  In addition, no Vainshten mechanism is needed to screen fifth forces produced by the scalar field. It is also worthwhile mentioning that recently developed positivity bounds that significantly shrink the viable parameter space of the Lorentz-invariant massive gravity theory \cite{Cheung:2016yqr,Bonifacio:2016wcb,Bellazzini:2017fep,deRham:2017xox} rely on Lorentz invariance of the theory and therefore do not apply to MTMG.

MTMG has two branches of stable cosmological solutions: a self-accelerating branch and a normal branch, both of which can accommodate the same cosmological background evolution as the $\Lambda$CDM in GR. Differences from GR show up at linear perturbation level (and above). In the self-accelerating branch, the scalar and vector perturbations have an identical phenomenology to that of $\Lambda$CDM in GR for any values of the graviton mass, while the massive tensor modes possess a nonzero mass, which is in turn responsible for the late-time acceleration of our universe at the background level. The normal branch, on the other hand, has differing dynamics for the scalar sector as well, which can produce differences in evolution at late times compared to GR (or the self-accelerating branch of MTMG). These differences can be identified though linear-perturbation observables such as $G_{\text{eff}}$ as well as the  integrated Sachs-Wolfe (ISW) effect, and therefore can produce different predictions to compare to cosmological data. 

In this paper we expand on the work done in \cite{DeFelice:2016ufg} by, not only examining RSD data, but also considering the fit of the normal branch of MTMG to data that describe the cross-correlation between galaxy overdensity perturbations and the change of the ISW perturbation field.  We aim to ascertain whether the normal branch of MTMG can reduce the tension between early and late cosmological data, including the tension in the measurement of $\sigma_8$ (which is related to the apparent need of weak gravity in RSD data), and whether it  survives this ISW-galaxy cross-correlation bound. To this end, we run a Chi-squared fit of the ISW-galaxy power cross-correlation $C_l^{\text{ISW}g}$ which is especially effective at bringing to light effects due to different dark energy or modified gravity models.  The ISW effect is produced from photons passing through time varying gravitational potential wells. While most of the CMB anisotropies are generated at early times on the last scattering surface, the difference in energy that photons gain or lose by passing through a time dependent potential well, can also produce new fluctuations.  During dark matter domination gravitational potential wells do not evolve; however, once dark energy\footnote{From CMB experiments, constraints on curvature are so strong that it is reasonable to assume the main effect of ISW comes from dark energy.} starts to dominate, potential wells can time-evolve. This effect can therefore be a powerful probe of the properties of dark energy or modified gravity.  Measuring the ISW effect alone however, is challenging since its signal is so much weaker than the rest of the CMB anisotropies. Moreover, the most significant effects of ISW reside at large scales which also coincide with the most problematic influence of cosmic variance. To address this issue, examining the cross-correlation between the temperature ISW and density of galaxy measurements can enhance our ability to trace the potential wells that cause the anisotropies \cite{Afshordi:2003xu,Afshordi:2004kz,Pogosian:2005ez,Corasaniti:2005pq,Giannantonio:2008zi,Kimura:2011td,Tommaso:2012,Giannantonio:2013kqa,Ade:2015xua}.

As we will see later on, the properties of the normal branch of MTMG are indeed emphasized by the ISW effect. We evaluate the combined Chi-squared for three different data sets (RSD data, 2d-mass data and SDSS ISW data), and put constraints on the free extra parameter $\theta$ of the simplest MTMG theory in the normal branch. The ISW-galaxy cross-correlation can put bounds on the free parameter of this branch of the theory, $\theta$. As RSD and ISW-galaxy cross-correlation data become more precise, i.e.\ the error bars shrink, it should be possible to distinguish $\Lambda$CDM in GR (or the self-accelerating branch of MTMG) from the normal branch of MTMG. At the moment ISW-galaxy cross-correlation data do not exclude the normal branch of MTMG at all, and in fact, if we assume the Planck best-fit values for the background parameters (namely $H_0$ and $\Omega_{m0}$), the normal branch of MTMG seems to be a better fit to the data we considered. 

%\begin{itemize}
%\item recent measurements of $H_0$ of Planck and SN seem in tension
%\item $\Lambda$CDM measurements of growth of structure seem higher then RSD measurements (as in other paper)
%\item MTMG can help explain, relieve this tension
%\item Describe MTMG properties and advantages compared to dRGT and other massive gravity
%\item descrive the two branches and how the normal branch can help with this problem
%\item How looking at ISWg measurement can help
%\end{itemize}

%\begin{enumerate}
%\item Massive gravities: usually 5 modes and Lorentz invarient, but suffers from Boulware-Deser ghost and instability on FRW.

%\item MTMG: same homogeneous and isotropic background cosmology and linear tensor perturbations as dRGT but no scalar and vector degrees of freedom at non linear level. \item 2 massive tensor modes, by breaking Lorenz symmetry 
%\item makes FRW, dS background stable
%\item two solution branches: 1) self accelerating branch which imitates GR and the graviton mass takes the role of the cosmological constant/ 2) normal branch which is the same as GR at high redshifts, but with different matter perturbation evolution at low redshifts. 
%\end{enumerate}

\section{The model}

In this section we quickly review how MTMG is defined, and more details can be found in \cite{DeFelice:2015hla,DeFelice:2015moy}. In order to build up the Lagrangian of the theory we first need to define a three-dimensional fiducial metric,
\begin{equation}
\tilde\gamma^{ij} = \delta_{IJ} E^I{}_i E^J{}_j\, 
\end{equation}
where $E^I{}_i$ is the fiducial three-vielbein, as well as a second three-tensor
\begin{equation}
\tilde{\zeta}^i{}_j =\frac1M E_L{}^i {\dot{E}}^L{}_j\,.
\end{equation}

Both these fields are considered to be given by external fields (as long as we make use of the unitary
gauge for the St\"uckelberg fields). In addition to these fields we need the three-dimensional
physical metric
\begin{equation}
\gamma_{ij} = \delta_{IJ} e^I{}_i\, e^J{}_j \, ,
\end{equation}
where $e^I{}_i$ is the physical three-dimensional vielbein. After having introduced ${\tilde\gamma}_{ij}$, ${\tilde\zeta}^i{}_j$, and $\gamma_{ij}$, we can define
\begin{equation}
\kappa^m{}_l\,\kappa^l{}_n = {\tilde\gamma}^{ms}\gamma_{sn}\,,\qquad k^m{}_j \kappa^j{}_n = \delta^m{}_n\, ,
\end{equation}
so that $k^m{}_j$ is the inverse tensor of $\kappa^j{}_n$.

Out of these quantities, by calling $\kappa=\kappa^i{}_i$, we introduce the following three-tensor
\begin{equation}
\Theta^{ij} = \frac{\sqrt{\tilde\gamma}}{\sqrt\gamma}
\{c_1 (\gamma^{il}\kappa^j{}_l + \gamma^{jl} \kappa^i{}_l)
+ c_2 [\kappa (\gamma^{il}\kappa^j{}_l + \gamma^{jl}\kappa^i{}_l) - 2{\tilde\gamma}^{ij} ]\} + 2c_3\gamma^{ij}\,,
\end{equation}
which is proportional to the difference between the values of the two canonical momenta $\pi^{ij}$ between
this theory and GR.

Having defined all the building blocks of the theory we can write down the action of MTMG in the metric formalism as follows
\begin{eqnarray}
S &=& \frac{\Mpl^2}2 \int d^4x N \sqrt\gamma 
[{}^{(3)}R + K^{ij}K_{ij} - K^2]
+ \frac{\Mpl^2}2\sum_{i=1}^4 \int d^4x \mathcal{L}_i\nonumber\\
&&{}+\frac{\Mpl^2}2 \int d^4x N \sqrt{\gamma} \left(\frac{m^2N\lambda}{4N}\right)^2\left(\Theta^{ij} \Theta_{ij} -\frac12\,\Theta^2\right)\nonumber\\
&&{}- \frac{\Mpl^2}2\int d^4x\sqrt{\gamma}[\lambda\mathcal{C}_0-(D_n{\lambda}^i)\mathcal{C}^n{}_i]+S_{\rm matter}\,,\\
\mathcal{L}_1 &=& -m^2 c_1 \tilde{a}^3 (N + M\kappa)\,,\\
\mathcal{L}_2 &=& -\frac12\,m^2 c_2 \tilde{a}^3 (2N\kappa + M\kappa^2 - M \kappa^i{}_j\,\kappa^j{}_i)\,,\\
\mathcal{L}_3 &=& -m^2 c_3 \sqrt\gamma (M + N k)\,,\\
\mathcal{L}_4 &=& -m^2 c_4 \sqrt\gamma N\,,\\
\mathcal{C}_0 &=& \frac12\,m^2\,K_{ij}\Theta^{ij}
- m^2M\left\{\frac{\sqrt{\tilde\gamma}}{\sqrt{\gamma}} [c_1\tilde\zeta 
+ c_2(\kappa\tilde\zeta - \kappa^m{}_n{\tilde\zeta}^n{}_m)]
+ c_3 k^m{}_n {\tilde\zeta}^n{}_m\right\},\\
\mathcal{C}^n{}_i &=& -m^2M\left\{\frac{\sqrt{\tilde\gamma}}{\sqrt{\gamma}}\left[\frac12\,(c_1+c_2\kappa)(\kappa^n{}_i+\gamma^{nm}\kappa^l{}_m\gamma_{li})
-c_2\kappa^n{}_l\kappa^l{}_i\right]+c_3\delta^m{}_i\right\}.
\end{eqnarray}
where $N$ is the lapse function, $k=k^i{}_i$, $\Theta=\Theta^{ij}\gamma_{ij}$, $\tilde\zeta={\tilde\zeta}^i{}_i$, $K_{ij}$ is the extrinsic curvature ($K$ being its trace), and $D_n$ represents the covariant derivative with respect to the physical 3D metric, $\gamma_{ij}$. 
Here, we have introduced two Lagrange multipliers, one scalar $\lambda$, and one three-vector $\lambda^i$ in order to implement four constraints. These constraints are necessary to eliminate the
scalar and vector perturbations of dRGT massive gravity without spoiling its self-accelerating FLRW background
solutions. Notice also that we used the unitary gauge, so that  $M = M(t)$, and $E^I{}_j = \tilde a(t)\, \delta^I{}_j$. Finally, we also introduced the matter fields in terms of a perfect fluid, which, in what follows, will be considered to be a pressureless dust fluid.

It can be shown that for this theory the Friedmann equation, can be written in terms of the Hubble factor $H = \dot{a}/(aN)$ and $X \equiv \tilde{a}/a$ (where $a$ is the scale factor of the physical metric and $\tilde{a}$ is the fiducial scale factor that corresponds to ${\tilde\gamma}^{ij}$) as follows
\begin{equation}
3\Mpl^2 H^2 = \frac{\Mpl^2m^2}2 (c_4 + 3c_3 X + 3c_2X^2 + c_1X^3)+\rho_m\,.
\end{equation}
Furthermore, using this Friedmann equation, and varying the Lagrangian with respect to $\lambda$, one finds the following equation of motion which has two branches of possible solutions,
\begin{equation}
(c_3 + 2c_2\,X + c_1X^2) (\dot{X} + NHX - MH)=0 \,.\label{eq:branches}
\end{equation}
The self-accelerating branch, for which $c_3 + 2c_2X + c_1X^2 = 0$, has the property that $X={\rm const}$, so that even when the pure cosmological constant term (proportional to $c_4$) is absent, the Friedmann equation will, in general, still {\bf possess} a nonzero effective cosmological constant. For the self-accelerating branch, it can be shown, see \cite{DeFelice:2015moy}, that the dynamics of both scalar and vector modes exactly coincides with the one of GR.
In both branches the tensor modes acquire a nonzero mass given by
\begin{equation}
\mu^2 = \frac12\,m^2\, X [c_2 X + c_3 + r X (c_1X + c_2)] \,,
\end{equation}
where $r\equiv M/(NX)$. In the following we will define $\theta\equiv\mu^2/H_0^2$.  %\textcolor{blue}{To preserve stability one must require $\mu^2>0$. }

In the normal branch, i.e.\ the other solution of Eq.\ (\ref{eq:branches}), both the background and the scalar-perturbation fields have different dynamics from those of the self-accelerating branch. This is the case even in the simplest scenario for which $r=1$, i.e.\ $X={\rm const}$. In this paper, we will focus on this last case, the simplest implementation of the normal branch, together with the self-accelerating branch.

\section{ISW-galaxy cross-correlations}
In this section we give an overview of the ISW-galaxy cross-correlation. The ISW effect creates CMB anisotropies from photons traveling through time varying gravitational potential wells on their path toward us. The photons will gain energy by entering a well and lose energy by climbing out again. If the potential evolves in the time that the photon is passing through it, there will be a net gain or loss of energy, which will add to the CMB anisotropy spectrum in the form of temperature fluctuations. The ISW effect can therefore be used to map out the evolution of these gravitational wells and gives us important information about the dynamics, especially at lower redshifts. Since the ISW effect signal is  weaker than the other CMB anisotropies, it is effectual to look at the ISW-galaxy cross-correlation which helps enhance the signal.

In the following we will work in units for which $c=1$, so that
$H_{{\rm Standard\ Units}}=\frac{1}{a}\,\frac{da}{dt}=\frac{c}{a}\,\frac{da}{dx^{0}}=c\,H$ and therefore the Hubble constant has units of Mpc$^{-1}$:

 \begin{equation}
H_{0}=\frac{H_{{\rm 0,Standard\ Units}}}{c}\approx3.33\cdot10^{-4}h\,{\rm Mpc}^{-1}.
\end{equation}
%\begin{equation}
%H_{0}=\frac{H_{{\rm 0,Standard\ Units}}}{c}=\frac{100\,h\,{\rm Km/s/Mpc}}{299792.458\,{\rm Km/s}}\approx3.33\cdot10^{-4}h\,{\rm Mpc}^{-1}\,,
%\end{equation}
%so that the horizon measures
%\begin{equation}
%\frac{1}{H_{0}}\approx2998\,h^{-1}\,{\rm Mpc}\,.
%\end{equation}
We start by defining the metric perturbations as
\begin{equation}
ds^{2}=-(1+2\alpha)N^{2}dt^{2}+2N\partial_{i}\chi dt\,dx^{i}+a^{2}[(1+2\zeta)\delta_{ij}+2\partial_{i}\partial_{j}s]dx^{i}dx^{j}\,,
\end{equation}
and the gauge invariant fields as
\begin{eqnarray}
\Psi & = & \alpha+\frac{\dot{\chi}}{N}-\frac{1}{N}\,\frac{d}{dt}\left(\frac{a^{2}\dot{s}}{N}\right)\,,\\
\Phi & = & -\zeta-H\chi+a^{2}H\,\frac{\dot{s}}{N}\,,\\
\delta & = & \frac{\delta\rho}{\rho}+3\zeta\,,
\end{eqnarray}
where $\Psi$ and $\Phi$ are the Bardeen gravitational potentials and $\delta$ the matter perturbation. The choice of this $\delta$ is driven by the good behavior at low values of $k$.  Later on we will also discuss the behavior of the $\delta_m$ field defined by the other gauge invariant combination, namely $\delta_m+3\,H\,v_m$ where $v_m$ is the velocity perturbation of the dust fluid. The choice of $\delta$ is convenient in order to evaluate the integrals for the ISW observable. At higher values of $k$, it does not matter which gauge we choose in order to study $\delta\rho/\rho$, i.e.\ $\delta\approx\delta_m$.

MTMG has additional constraints, compared to dRGT theory, which are necessary to reduce the propagating degrees of freedom. One of these constraints, in the context of cosmological linear perturbation theory, leads to $\zeta=0$. In this case, the gauge invariant field defined above, $\delta$, reduces to $\delta=\delta\rho/\rho$.

We can write the ISW contribution to the temperature perturbation in terms of the field $\psi_{{\rm ISW}}\equiv\Psi+\Phi$,
\begin{equation}
\frac{\Delta T_{{\rm ISW}}(\eta,\hat{\bm{\theta}})}{T}  =  \lim_{\eta\to\eta_{0}}\int_{\eta_{r}}^{\eta}d\eta\,\frac{\partial\psi_{{\rm ISW}}}{\partial\eta}= -\lim_{z\to0}\int_{z}^{z_{r}}dz\,\frac{\partial\psi_{{\rm ISW}}}{\partial z}\,.
\end{equation}
%\begin{eqnarray}
%\frac{\Delta T_{{\rm ISW}}(\eta,\hat{\bm{\theta}})}{T} & = & \lim_{\eta\to\eta_{0}}\int_{\eta_{r}}^{\eta}d\eta\,\frac{\partial\psi_{{\rm ISW}}}{\partial\eta}=\lim_{\eta\to\eta_{0}}\int_{\eta_{r}}^{\eta}d\eta\,\frac{\partial}%{\partial\eta}[\Psi+\Phi]\,,\nonumber \\
% & = & \lim_{z\to0}\int_{z_{r}}^{z}dz\,\frac{\partial\psi_{{\rm ISW}}}{\partial %z}=-\lim_{z\to0}\int_{z}^{z_{r}}dz\,\frac{\partial\psi_{{\rm ISW}}}{\partial z}\,.
%5\end{eqnarray}
where $\eta_{r}$ corresponds to conformal time at recombination and $z_r$ the corresponding redshift of recombination.
We will assume that matter perturbation  $\delta(\bm{k},a)$ can be broken up in $k$  (with $k$ here and in the following representing the wave vector of the perturbations written in Fourier space) and time dependent parts. This is typically
true if the equation of motion for $\delta$ depends only mildly on $k$ so that we can write
\begin{equation}
\delta(\bm{k},a)  =  D(a)\,f(\bm{k})=\frac{D(a)}{D(z=0)}\,D_{0}(z=0)\,f(\bm{k})=D\,\frac{\delta(0,\bm{k})}{D_{0}}
\end{equation}
and its derivative with respect to the $e$-fold number $\mathcal{N}$, 
\begin{equation}
\frac{d\delta}{d\mathcal{N}}  =  \frac{dD}{d\mathcal{N}}\,\frac{\delta(0,\bm{k})}{D_{0}}\,.
\end{equation}
The $e$-fold number and its derivative are related to the scale factor and redshift by $\mathcal{N} =\ln a = -\ln(1+z)$.
%Later on we will consider the field
%\begin{eqnarray}
%\frac{\partial\psi_{{\rm ISW}}}{\partial\mathcal{N}} & = & Q_{\psi}\,\delta+P_{\psi}\,\frac{d\delta}{d\mathcal{N}}\,,
%\end{eqnarray}
%where $Q_{\psi}$ and $P_{\psi}$ are known functions yet to be introduced. In this case we can write
%\begin{eqnarray}
%\frac{\partial\psi_{{\rm ISW}}}{\partial\mathcal{N}} & = & Z_{{\rm ISW}}\,\frac{\delta(0,\bm{k})}{D_{0}}\,.\\
%Z_{{\rm ISW}} & \equiv & Q_{\psi}D+P_{\psi}\,\frac{dD}{d\mathcal{N}}\,,
%\end{eqnarray}
%where $D_{0}=D(\mathcal{N}=0)$. The expression for $Z_{{\rm ISW}}$will
%be given later on.
Next let us consider the observed projected galaxy overdensity $g$, 
\begin{equation}
g=\int_{0}^{z_{{\rm rec}}}dz\,b(z)\,\phi(z)\,\delta(z,\chi,\hat{n})\,.
\end{equation}
where $b(z)$ is the redshift dependent bias and $\phi(z)$ is the window function of the form 
\begin{equation}
\phi(z)=\frac{\beta}{\Gamma[(m+1)/\beta]\,z_{0}}\left(\frac{z}{z_{0}}\right)^{m}e^{-(z/z_{0})^{\beta}}\,,
\end{equation}
for which
\begin{equation}
\int_{0}^{\infty}\phi(z)\,dz=1\,.
\end{equation}
Here we choose to use a redshift dependent window function and bias because we want to try to find constraints which are as bias independent as possible. In particular we will only select ISW experiments which have a window function that is peaked around a particular experiment-dependent redshift. In this case our results will not depend on possible time variations of the bias itself.
The ISW-galaxy cross-correlation amplitude $C_{l}^{\rm GI}$, can be found by evaluating the two-point function  
\begin{equation}
C_{l}^{\rm GI}=\left\langle a_{lm}^{{\rm ISW}}a_{lm}^{G*}\right\rangle ,
\end{equation}
where we have expanded both the ISW and Galaxy overdensity integrals with spherical harmonics in the following way
\[
X(\hat{n})=\int_{0}^{z_{{\rm \infty}}}dzX(z,\chi(z)\,\hat{n})=\sum_{l,m}a_{lm}^{X}\,Y_{lm}(\hat{n})
\]
and we find, by using $Y_{lm}^{*}(\hat{n})=(-1)^{m}\,Y_{l,-m}(\hat{n})$, the spherical harmonic coefficients are,
\begin{eqnarray}
a_{lm}^{X} & = & \int d^{2}nY_{lm}^{*}(\hat{n})X(\hat{n})\nonumber \\
 %& = & \int d^{2}n\int dzY_{lm}^{*}(\hat{n})X(z,\bm{x})\nonumber \\
 %& = & \int d^{2}n\,\int dz\,\int\frac{d^{3}k}{(2\pi)^{3}}\,Y_{lm}^{*}(\hat{n})\,X(z,\bm{k})e^{-i\bm{k}\cdot\bm{x}}\nonumber \\
 %& = & \int d^{2}n\,dz\,\frac{d^{3}k}{(2\pi)^{3}}\,Y_{lm}^{*}(\hat{n})\,X(z,-\bm{k})e^{i\bm{k}\cdot\bm{x}}\nonumber \\
 %& = & 4\pi\int d^{2}n\,dz\,\frac{d^{3}k}{(2\pi)^{3}}\,Y_{lm}^{*}(\hat{n})\,X(z,-\bm{k})\sum_{L}\sum_{M}(i)^{L}Y_{LM}(\hat{n})^{*}Y_{LM}(\hat{k})j_{L}(k\chi(z))\nonumber \\
 %& = & \int dz\,\frac{d^{3}k}{2\pi^{2}}\,X(z,-\bm{k})\sum_{L}\sum_{M}(i)^{L}Y_{LM}(\hat{k})j_{L}(k\chi)\int d^{2}n\,Y_{lm}^{*}(\hat{n})Y_{LM}(\hat{n})^{*}\nonumber \\
 %& = & \int dz\,\frac{d^{3}k}{2\pi^{2}}\,X(z,-\bm{k})\sum_{L}\sum_{M}(i)^{L}Y_{LM}(\hat{k})j_{L}(k\chi)(-1)^{M}\int d^{2}n\,Y_{lm}^{*}(\hat{n})Y_{L,-M}(\hat{n})\nonumber \\
 %& = & \int dz\,\frac{d^{3}k}{2\pi^{2}}\,X(z,-\bm{k})\sum_{L}\sum_{M}(i)^{L}Y_{LM}(\hat{k})(-1)^{M}j_{L}(k\chi)\delta_{Ll}\delta_{M,-m}\nonumber \\
 %& = & \frac{1}{2\pi^{2}}\,(i)^{l}\int dz\,d^{3}k\,X(z,-\bm{k})Y_{l,-m}(\hat{k})(-1)^{-m}j_{l}(k\chi)\nonumber \\
 & = & \frac{1}{2\pi^{2}}\,(i)^{l}\int dz\,d^{3}k\,X(z,-\bm{k})Y_{lm}(\hat{k})^{*}j_{l}(k\chi)\,.
\end{eqnarray}
Using these we can then find the ISW-galaxy cross-correlation amplitude to be,
\begin{eqnarray}
C_{l}^{\rm GI} & = &\frac{1}{2\pi^{2}} \left\langle [(i)(-i)]^{l}\int_{z_{0}}^{z_{i}}dz_{1}\,d^{3}k_{1}\,[-\partial_{z_{1}}\psi_{{\rm ISW}}(-\bm{k}_{1},z_{1})]Y_{lm}^{*}(\hat{k}_{1})j_{l}(k_{1}\chi_{1})\right.\nonumber \\
&& {}\times\left.\int_{z_{0}}^{z_{i}}dz_{2}\,d^{3}k_{2}\,\phi(z_{2})b_{s}\delta(-\bm{k}_{2},z_{2})^{*}Y_{lm}(\hat{k}_{2})j_{l}(k_{2}\chi_{2})\right\rangle \nonumber \\
 & = & \frac{2}{\pi D_{0}^{2}}\int_{k_{m}}^{k_{M}}k^{2}dk\,P(k)\int_{\mathcal{N}_{0}}^{\mathcal{N}_{i}}d\mathcal{N}_{1}\,j_{l}(k\chi_{1})Z_{{\rm ISW}}(\mathcal{N}_{1})\int_{\mathcal{N}_{0}}^{\mathcal{N}_{i}}d\mathcal{N}_{2}e^{-\mathcal{N}_{2}}\,\phi(\mathcal{N}_{2})\,b_{s}\,D(\mathcal{N}_{2})\,j_{l}(k\chi_{2})\,.\label{eq:CIG}
\end{eqnarray}
Here $j_l$ are spherical bessel functions and $Z_{\rm ISW}(\mathcal{N})$ is related to the derivative of the ISW field,
\begin{equation}
\frac{\partial\psi_{{\rm ISW}}}{\partial\mathcal{N}}  =  Z_{{\rm ISW}}\,\frac{\delta(0,\bm{k})}{D_{0}}\,
\end{equation}
where $D_{0}=D(\mathcal{N}=0)$. In the end, we find that the ISW observable does not depend on the choice of $D$ to be the amplitude of $\delta$ or $\delta_m$. The expression for $Z_{{\rm ISW}}$ is instead given explicitly in Eq.\ (\ref{eq:evDpsiISW}).
We can write
\begin{equation}
\chi=\int_{\eta}^{\eta_{0}}d\eta=\int_{z}^{0}\frac{d\eta}{Ndt}\,\frac{Ndt}{da}\frac{da}{dz}\,dz=\int_{0}^{z}\frac{dz}{H}=-\int_{0}^{\mathcal{N}}\frac{d\mathcal{N}e^{-\mathcal{N}}}{H},
\end{equation}
and define the dimensionless quantity
\begin{eqnarray}
\bar{\chi} & = & H_{0}\chi=\int_{0}^{z}\frac{H_{0}}{H}dz.
\end{eqnarray}

The galaxy-galaxy correlation is calculated in the same way and will be necessary for our determination of the bias as shown in Sec.~\ref{sec:bias}. Following the same procedure we used to find Eq.\ (\ref{eq:CIG}), we find the galaxy-galaxy cross-correlation as
\begin{eqnarray}
C_{l}^{\rm GG}=\left\langle a_{lm}^{\rm G}a_{lm}^{\rm G*}\right\rangle  & = & \left\langle [(i)(-i)]^{l}\frac{1}{2\pi^{2}}\int_{z_{0}}^{z_{i}}dz_{1}\,d^{3}k_{1}\,\phi(z_{1})b_{s}\delta(-\bm{k}_{1},z_{1})Y_{lm}(\hat{k}_{1})^{*}j_{l}(k_{1}\chi_{1})\right.\nonumber \\
 &  & {}\times\left.\frac{1}{2\pi^{2}}\int_{z_{0}}^{z_{i}}dz_{2}\,d^{3}k_{2}\,\phi(z_{2})b_{s}\delta(-\bm{k}_{2},z_{2})^{*}Y_{lm}(\hat{k}_{2})j_{l}(k_{2}\chi_{2})\right\rangle  \nonumber \\
 & = & \frac{2}{\pi D_{0}^{2}}\int_{k_{m}}^{k_{M}}dk\,k^{2}P(k)\int_{\mathcal{N}_{i}}^{\mathcal{N}_{0}}d\mathcal{N}_{1}\,e^{-\mathcal{N}_{1}}\,\phi(\mathcal{N}_{1})\,b_{s}\,D(\mathcal{N}_{1})\,j_{l}(k\chi_{1})\nonumber \\
 &  & {}\times\int_{\mathcal{N}_{i}}^{\mathcal{N}_{0}}d\mathcal{N}_{2}\,e^{-\mathcal{N}_{2}}\,\phi(\mathcal{N}_{2})b_{s}\,D(\mathcal{N}_{2})j_{l}(k\chi_{2})\,.\label{eq:CGG}
\end{eqnarray}
\subsection{Background evolution}
The evolution of the cosmological background depends on the equations of motion for $\Omega_m\equiv\rho_m/(3\Mpl^2 H^2)$, $Y\equiv H_0/H$ and $\bar{\chi}$, which can be written as
%\begin{equation}
%\frac{d\bar{\chi}}{dz}=\frac{H_{0}}{H}=\sqrt{Y},
%\end{equation}
%and
\begin{eqnarray}
\frac{d\bar{\chi}}{d\mathcal{N}} & = & -e^{-\mathcal{N}}\sqrt{Y}=-\left(\frac{\Omega_{m}\sqrt{Y}}{\Omega_{m0}}\right)^{1/3}\,,\\
\frac{d\Omega_{m}}{d\mathcal{N}} & = & -3\Omega_{m}(1-\Omega_{m})\,,\\
\frac{dY}{d\mathcal{N}} & = & 3\,Y\,\Omega_{m}\,.
\end{eqnarray}
with initial conditions $Y(\mathcal{N}=0)=1$, $\bar{\chi}(\mathcal{N}=0)=0$, and $\Omega_{m}(\mathcal{N}=0)=\Omega_{m0}$, where this last value is obtained by the Planck best-fit cosmological parameters. Here we have neglected the contribution of radiation as the data we are going to analyze only depend on the dust and dark-energy components. In this case we will set the initial $e$-folding number $\mathcal{N}_i=-6$, i.e.\ in the deep matter era.
 
Finally, the matter power spectrum is defined as 
\begin{equation}
P=2\pi^{2}\,\delta_{H}^{2}\,[T_{m}(k)]^{2}\,\left(\frac{k}{H_{0}}\right)^{n_{s}}H_{0}^{-3}
\end{equation}
where $H_{0}$, is the Hubble parameter in units Mpc$^{-1}$ and we fix $k$ to have dimensions of Mpc$^{-1}$. The transfer function $T_{m}$ is evaluated by using the fitting formula given by Eisenstein and Hu \cite{Eisenstein:1997ik,Eisenstein:1997jh}.

\section{Cross-correlations via ODE solver}

Let us consider once more the triple integrals of $C_{l}^{{\rm GI},\alpha}$ and $C_{l}^{{\rm GG},\alpha}$ that  we are investigating. In particular we can write them as
%\begin{eqnarray}
%C_{l}^{{\rm GI}} & = & \frac{4\pi\delta_{H}^{2}b_{s}}{D_{0}^{2}}\int_{\ln k_{m}}^{\ln k_{M}}d\ln k\,[T_{m}(k)]^{2}\left(\frac{k}{H_{0}}\right)^{n_{s}+3}\int_{\mathcal{N}_{i}}^{\mathcal{N}_{0}}d\mathcal{N}_{1}\,j_{l}(k\chi_{1})Z_{{\rm ISW}}\int_{\mathcal{N}_{i}}^{\mathcal{N}_{0}}d\mathcal{N}_{2}e^{-\mathcal{N}_{2}}\,\phi(\mathcal{N}_{2})\,Dj_{l}(k\chi_{2})\,,\\
%C_{l}^{{\rm GG}} & = & \frac{4\pi\delta_{H}^{2}b_{s}^{2}}{D_{0}^{2}}\int_{\ln k_{m}}^{\ln k_{M}}d\ln k\,[T_{m}(k)]^{2}\left(\frac{k}{H_{0}}\right)^{n_{s}+3}\left[\int_{\mathcal{N}_{i}}^{\mathcal{N}_{0}}d\mathcal{N}_{1}\,e^{-\mathcal{N}_{1}}\,\phi(\mathcal{N}_{1})\,D(\mathcal{N}_{1})\,j_{l}(k\chi_{1})\right]^{2}.
%\end{eqnarray}
%Therefore we can rewrite the integrals we need to find as
\begin{eqnarray}
C_{l}^{{\rm GI},\alpha} & = & 4\pi\,\bar{\delta}_{H}^{2}b_{s}^{\alpha}\int_{\ln(k_{m})}^{\ln(k_{M})}\,d(\ln k)[T_{m}(k)]^{2}\left(\frac{k}{H_{0}}\right)^{n_{s}+3}F_{G}^{\alpha}(l,\theta,k/H_{0})\,F_{I}(l,\theta,k/H_{0})\,,\\
C_{l}^{{\rm GG},\alpha} & = & 4\pi\,\bar{\delta}_{H}^{2}(b_{s}^{\alpha})^{2}\int_{\ln(k_{m})}^{\ln(k_{M})}\,d(\ln k)[T_{m}(k)]^{2}\left(\frac{k}{H_{0}}\right)^{n_{s}+3}[F_{G}^{\alpha}(l,\theta,k/H_{0})]^{2}\,,
\end{eqnarray}
where
\begin{eqnarray}
\label{F1}
F_{G}^{\alpha}(l,\theta,k/H_{0}) & = & \lim_{\mathcal{N}\to0}\int_{\mathcal{N}_{i}}^{\mathcal{N}}d\mathcal{N}_{2}\,e^{-\mathcal{N}_{2}}\phi^{\alpha}(\mathcal{N}_{2})D(\mathcal{N}_{2})\,j_{l}[(k/H_{0})\bar{\chi}(\mathcal{N}_{2})]\,,\\
\label{F2}
F_{I}(l,\theta,k/H_{0}) & = & \lim_{\mathcal{N}\to0}\int_{\mathcal{N}_{i}}^{\mathcal{N}}d\mathcal{N}_{1}\,Z_{{\rm ISW}}\,j_{l}[(k/H_{0})\bar{\chi}(\mathcal{N}_{1})]\,.
\end{eqnarray}
It should be noted that the quantity $F_{I}$ does not depend on the experimental window function, but only on the behavior of the
ISW evolution. The bias function $b_{s}^{\alpha}$
can in theory change in time, but here it is assumed to be constant inside the
rather restricted $z$-interval allowed by the window functions, $\phi^{\alpha}(z)$. The  label $\alpha$ on the window functions stands for the two chosen ISW experiments (2d-mass and SDSS data)  which were taken by the following \cite{Giannantonio:2008zi}. 
By calling $k=e^{\sigma}$/Mpc, we solve the following system of ordinary differential equations (ODE),
\begin{eqnarray}
\frac{dk}{d\sigma} & = & k\,,\\
\frac{dY_{l}^{{\rm GG}\alpha}}{d\sigma} & = & [T_{m}(k)]^{2}\left(\frac{k}{H_{0}}\right)^{n_{s}+3}[F_{G}^{\alpha}(l,\theta,k/H_{0})]^{2}\,,\\
\frac{dY_{l}^{{\rm GI}\alpha}}{d\sigma} & = & [T_{m}(k)]^{2}\left(\frac{k}{H_{0}}\right)^{n_{s}+3}F_{G}^{\alpha}(l,\theta,k/H_{0})\,F_{I}(l,\theta,k/H_{0})\,,\\
k(\sigma_{i}) & = & k_{m}=\exp(\sigma_{i})\,,\\
Y_{l}^{{\rm GG},\alpha}(\sigma_{i}) & = & 0\,,\\
Y_{l}^{{\rm GI},\alpha}(\sigma_{i}) & = & 0\,,\\
C_{l}^{{\rm GI},\alpha} & = & 4\pi\,\bar{\delta}_{H}^{2}\,b_{s}^{\alpha}\,Y_{l}^{{\rm GI},\alpha}\,,\\
C_{l}^{{\rm GG},\alpha} & = & 4\pi\,\bar{\delta}_{H}^{2}(b_{s}^{\alpha})^{2}\,Y_{l}^{{\rm GG},\alpha}\,,\label{eq:ClGG}
\end{eqnarray}
up to $\sigma_{f}=\ln k_{M}$. Let us now discuss the integration
limits. Although the integral should in principle be performed from
$-\infty<\sigma<\infty$, very negative values of $\sigma$ would correspond
to selecting superhorizon modes, and very large and positive values
of $\sigma$ would be affected by nonlinearities. We have therefore selected the
range of $\sigma$ such that the triple integral consistently reduces to the Limber
approximation for high values of $l$, and, further, does not change considerably
its value by extending the extrema of the integration interval. Finally these integrals depend on the free parameter $\theta$, which is proportional to the mass of the graviton ($\theta=0$ corresponds to $\Lambda$CDM). We fix all the other parameters to their Planck best-fit values.

To calculate the functions of Eq.\ref{F1} and \ref{F2}, we also need to  integrate the following ODE system:
\begin{eqnarray}
\frac{d\Omega_{m}}{d\mathcal{N}} & = & -3\Omega_{m}(1-\Omega_{m})\,,\\
\frac{dY}{d\mathcal{N}} & = & 3Y\Omega_{m}\,,\\
\frac{d\bar{\chi}}{d\mathcal{N}} & = & -\left(\frac{\Omega_{m}\sqrt{Y}}{\Omega_{m0}}\right)^{1/3}\,,\\
\frac{dD}{d\mathcal{N}} & = & \pi_{D}\,,\\
\frac{d\pi_{D}}{d\mathcal{N}} & = & -T_{2}\,\pi_{D}-T_{3}\,D,\\
\frac{dF_{G}^{\alpha}}{d\mathcal{N}} & = & \left(\frac{\Omega_{m}}{\Omega_{m0}Y}\right)^{1/3}\phi^{\alpha}(\mathcal{N})\,D(\mathcal{N})\,j_{l}[(k/H_{0})\bar{\chi}(\mathcal{N})]\,,\\
\frac{dF_{I}}{d\mathcal{N}} & = & Z_{{\rm ISW}}\,j_{l}[(k/H_{0})\bar{\chi}(\mathcal{N})]
\end{eqnarray}
where we have used the relation $e^{-3\mathcal{N}}=\Omega_{m}/(\Omega_{m0}\,Y)$.
It should be noted that the initial conditions for the background quantities $Y$, $\Omega_{m}$, and $\bar{\chi}$ are known only for today (1, $\Omega_{m0}$, 0, respectively),
and therefore we need to integrate backwards (only once) the first three ODEs by themselves from $\mathcal{N}=0$ to $\mathcal{N}=\mathcal{N}_{i}\equiv-6$,
i.e.\ deep in the matter era, in order to find $Y_{i}$, $\Omega_{mi}$, $\bar{\chi}_{i}$. The initial values for each function are then
\begin{eqnarray}
\Omega_{m}(\mathcal{N}_{i}) & = & \Omega_{mi}\,,\\
Y(\mathcal{N}_{i}) & = & Y_{i}\,,\\
\bar{\chi}(\mathcal{N}_{i}) & = & \bar{\chi}_{i}\,,\\
D(\mathcal{N}_{i}) & = & \delta_i\,,\\
\pi_{D}(\mathcal{N}_{i}) & = & \delta'_{i}\,,\\
F_{G}^{\alpha}(\mathcal{N}_{i}) & = & 0\,,\\
F_{I}(\mathcal{N}_{i}) & = & 0\,,
\end{eqnarray}
where the values of $\delta_i$ and $\delta'_{i}$ are found by imposing the two conditions $\delta_m(\mathcal{N}_i)=\exp(\mathcal{N}_i)$ and $\psi'_{{\rm ISW}}(\mathcal{N}_{i})=0$. Here, we recall that $\delta_m=\delta\rho/\rho+3Hv_m$ is the gauge invariant matter density fluctuation. The first condition corresponds to a normalization condition for the dust matter field, whereas the second one imposes the initial value for the integrand to vanish during the deep dust domination era.

We would like to take a moment to  explain why we have used this method to evaluate the triple integral. One way to solve a triple integral would be to implement a three-dimensional Simpson rule. Indeed this is a viable method; however, in order to reach the desired accuracy, one would need to increase the number of points in the 3D grid. This method therefore typically ends up being either quite slow or not very precise. On the other hand, in this work we use a method with two ODEs, one nested in the other, namely one for the $\mathcal{N}$-integral and the other one for the $k$-integral. We find that the explicit adaptive Runge-Kutta methods (respectively Dormand-Prince of  the eighth-order for the $\mathcal{N}$ integration; Cash-Karp of the fifht order for the $k$-integration) are very efficient to solve both the ODEs.  Compared to the 3D Simpson rule, this code is considerably faster and more precise. 

\subsection{The galaxy bias}\label{sec:bias}

In order to evaluate the ISW-galaxy cross-correlation we will need to determine the bias $b_s$. 
We reproduce the data $C_{l}^{{\rm GG}}$ fitted by a WMAP model using the best-fit $\Lambda$CDM parameter, and the
known values of the bias for such an experiment. In order not to suffer much from possible bias-time dependence  we selected those experiments, for which, the window function for the galaxies data
bins are considerably peaked in the redshift. We also consider the bias to be scale independent.
The experiments that meet these criteria are 2d mass  and SDSS (see, e.g.\ \cite{Giannantonio:2008zi}).  We
define $C_{l,{\rm WMAP}}^{{\rm GG}\alpha}$
as the values obtained by using Eq.\ (\ref{eq:ClGG}) with the
experimental values of the bias obtained assuming WMAP best-fit parameters \footnote{
In terms of the low-$l$ regime we are interested in, Planck data do not give appreciable improvements with respect to WMAP \cite{Giannantonio:2013kqa}}. For
any other value of the parameters, we find the bias by minimizing
the following chi-squared
\begin{equation}
\chi_{\alpha,{\rm bias}}^{2}\equiv\sum_{l=2}^{150}\bigl[C_{l,{\rm WMAP}}^{{\rm GG}\alpha}-4\pi\bar\delta_H^2(b_{s}^{\alpha})^{2}Y_{l}^{{\rm GG}\alpha}\bigr]^{2}\,.
\end{equation}
The $Y_{l}^{{\rm GG}\alpha}$ can be calculated for any value
of the parameters, whereas the $C_{l,{\rm WMAP}}^{GG,\alpha}$ corresponds
to the known data (which must be fitted). For $l>30$, we use the values
given by the Limber approximation (as we discuss further later on). Then, since
\begin{equation}
\frac{\partial\chi_{\alpha,{\rm bias}}^{2}}{\partial b_{s}^{\alpha}}=-8\pi\bar\delta_H^2 b_{s}^{\alpha}\Bigl\{\sum_{l}\bigl[C_{l,{\rm WMAP}}^{{\rm GG}\alpha}-4\pi\bar\delta_H^2(b_{s}^{\alpha})^{2}Y_{l}^{{\rm GG}\alpha}\bigr]Y_{l}^{{\rm GG}\alpha}\Bigr\}\,,
\end{equation}
we find that
\begin{equation}
b_{s}^{\alpha}=\sqrt{\frac{\sum_{l}C_{l,{\rm WMAP}}^{{\rm GG}\alpha}Y_{l}^{{\rm GG}\alpha}}{4\pi\bar\delta_H^2\sum_{l}(Y_{l}^{{\rm GG}\alpha})^{2}}}\,.
\end{equation}
We will adopt this method in order not to have free parameters other than $\theta$. We see that the bias, for different $\theta$-models changes only mildly (only by a few percent, well inside the typical error bars of the bias estimates, see e.g.\ \cite{Giannantonio:2008zi}) and so even keeping the same values for the bias of $\Lambda$CDM would not change the final results noticeably. This procedure is not new, it has been adopted before, and it is a way to fix the bias so that we can reduce as much as possible the number of free parameters \cite{Prat:2016xor,Biasrep}.

\subsection{On calculating $\delta_{H}$}
In order to calculate $\delta_{H}$, the amplitude of the power spectrum, we evaluate the following integral.
First we define the top-hat function
\begin{equation}
w_{{\rm TH}}(r,k)=\frac{3}{(k\,r)^{3}}\,[\sin(kr)-k\,r\,\cos(k\,r)]\,,
\end{equation}
and we evaluate the integral
\begin{equation}
I=\int_{\ln k_{m}}^{\ln k_{M}}d(\ln k)\left(\frac{k}{H_{0}}\right)^{n_{s}+3}\,[T_{m}(k)\,w_{{\rm TH}}(8/h,k)]^{2}\,.
\end{equation}
Then we find
\begin{equation}
\delta_{H}\equiv\frac{\sigma_{8}(0)}{\sqrt{I}}\,.
\end{equation}
where $w_{{\rm TH}}$ has been evaluated for the value of $r=8/h$ Mpc.
It should be noticed that $I$ does not depend on $\theta$. But $\sigma_{8}(\mathcal{N}=0)$
does depend on the value of $\theta$. In order to calculate the variable $I$, we can introduce the variable $k=e^{\sigma}$, and define
$I(\sigma)$ by replacing $\ln k_{M}$ with $\sigma$. So that
\begin{eqnarray}
\frac{dk}{d\sigma} & = & k\,,\\
\frac{dI}{d\sigma} & = & \left(\frac{k}{H_{0}}\right)^{n_{s}+3}\,[T_{m}(k)\,w_{{\rm TH}}(8/h,k)]^{2}\,,
\end{eqnarray}
and we integrate from $\sigma_{m}=\ln k_{m}$, at which $I(\sigma_{i})=0$,
and $k_{i}=k_{m}$, to $\sigma_{M}=\ln k_{M}$. Here $k_{m},k_{M}$
need to be chosen such that the integral does not change considerably
by extending the extrema of the integration interval.

Since 
\begin{equation}
\sigma_{8}(0)=\sigma_{8}(\mathcal{N}_{i})\,\frac{D_{0}}{D(\mathcal{N}_{i})}\,,
\end{equation}
then 
\begin{equation}
\frac{\delta_{H}}{D_{0}}\equiv\bar{\delta}_{H}=\frac{\sigma_{8}(\mathcal{N}_{i})}{D(\mathcal{N}_{i})}\,\frac{1}{\sqrt{I}}\,,
\end{equation}
which is also $\theta$ independent, as by assumption all the theories
will share the same $\sigma_{8}(N_{i})=\sigma_{8}^{{\rm \Lambda}{\rm CDM}}(N_{i})$ and $D(N_{i})\approx\exp(N_{i})$. Here we assume we get the normalization
at values of $k$ close to the peak of the power spectrum, say $k\approx0.1h$/Mpc.
In this case $K=k/H_0\approx300\gg1$, and we are in a high-$k$ linear regime for the perturbation variables. For these values of $k$, $\delta\rho/\rho\approx\delta\approx\delta_{m}$,
the value of the matter perturbation is gauge independent. In the
following we will be using the $\theta$-independent quantity, $\bar{\delta}_{H}$.

\section{The high momenta approximation}

In the high-$k$ limit, we can discuss the dynamics of the system
in a gauge-independent and clear way. In particular, in the normal branch of MTMG, by using the results given in \cite{DeFelice:2015moy}, we find
that in the subhorizon scales we can write:
\begin{equation}
\psi_{{\rm ISW}} = -\frac{3H_{0}^{2}\,\Omega_{m0}}{k^{2}}\,(1+z)\,\Sigma\,\delta\,,
\end{equation}
where we have introduced
\begin{equation}
\Sigma\equiv\frac{8-Y\theta(4+3\Omega_{m})}{2\,(2-Y\theta)^{2}}\,,
\end{equation}
and we have used the results of \cite{DeFelice:2016ufg}. At early times we have, as expected,
\begin{equation}
\lim_{Y\to0}\Sigma\to1\,.
\end{equation}
Therefore, in the normal branch of MTMG, we can write
\begin{equation}
\psi_{{\rm ISW}}(\bm{k},z)=-3\,H_{0}^{2}\,\Omega_{m0}\,(1+z)\,\Sigma\,\frac{D(z)}{D(z=0)}\,\frac{\delta(\bm{k},z=0)}{k^{2}}\,,
\end{equation}
so that
\begin{eqnarray}
\delta(k,z) & = & \delta(k)\,D(z)=\delta(k)\,D(z=0)\,\frac{D(z)}{D(z=0)}=\frac{D(z)}{D(z=0)}\,\delta(k,z=0)\,,\\
\psi_{{\rm ISW}}' & = & \Phi'+\Psi'=\frac{\partial}{\partial\eta}(\Phi+\Psi)=-\frac{3}{k^{2}}\,H_{0}^{2}\,\Omega_{m0}\,\frac{\partial}{\partial\eta}[(1+z)\,\Sigma(z)\,\delta(\bm{k},z)]\nonumber \\
 & = & -\frac{3}{k^{2}D(z=0)}\,H_{0}^{2}\,\Omega_{m0}\,\delta(\bm{k},z=0)\,\frac{\partial}{\partial\eta}[(1+z)\,\Sigma(z)\,D(z)]\,.
\end{eqnarray}

When this last quantity becomes negative at low redshift, then typically we will have anti-cross-correlation between ISW-galaxy cross-correlation.

\subsection{The Limber approximation}

Following the same motivation used for the high-$k$ approximation, it is also useful for large $l$,
to use the Limber approximation. First we note the following relation $\int dx\,\delta(f(x))=\sum_{i}\frac{\delta(x-a_{i})}{\left|\frac{df}{dx}(a_{i})\right|}$ where $f(a_{i})=0$,
and reiterate the definition of the $e$-folding number,

\begin{equation}
\mathcal{N}=\ln a\,,\qquad\frac{d\chi}{d\mathcal{N}}=-\frac{e^{-\mathcal{N}}}{H}\,.
\end{equation}
In this case for large $l$, we can make use of the Limber approximation,
namely
\begin{equation}
\int k^{2}dk\,j_{l}(k\chi_{1})j_{l}(k\chi_{2})F(k)\approx\frac{\pi}{2}\,\frac{\delta(\chi_{1}-\chi_{2})}{\chi_{1}^{2}}\,F(l_{12}/\chi_{1})\,,
\end{equation}
where $l_{12}=l+\frac{1}{2}$. 

We have taken the argument of the power spectrum to be the quantity
\begin{equation}
k=\frac{l_{12}}{\chi}=\frac{l_{12}H_{0}}{H_{0}\chi}=\frac{l_{12}H_{0}}{\bar{\chi}}\,,\qquad{\rm or}\qquad\frac{k}{H_{0}}\equiv K=\frac{l_{12}}{\bar{\chi}}\,.
\end{equation}
Focusing on high $l$, implies we  are also focusing on high $k$.
Therefore, we use the variables for the ODEs that correspond to dynamics valid for high-$k$, defined in
Eq.\ (\ref{eq:dynHighK}). Hence we have

\begin{equation}
 C_{l}^{{\rm GI}} =  \frac{2}{\pi}\left[\frac{3}{D(0)^{2}}\,H_{0}^{2}\,\Omega_{m0}\right]\int dz_{1}dz_{2}\,\left[\frac{\partial}{\partial z_{1}}[(1+z_{1})\,\Sigma(z_{1})\,D(z_{1})]\right]\phi(z_{2})D(z_{2})b_{s} \int k^{2}dk\,\frac{P(k)}{k^{2}}\,j_{l}(k\chi_{1})j_{l}(k\chi_{2})\,.
\end{equation}
This integral can then be rewritten as
\begin{eqnarray}
C_{l}^{{\rm GI},\alpha} & = & -\frac{2}{\pi}\left[\frac{3}{D(0)^{2}}\,\Omega_{m0}\right]\int_{\mathcal{N}_{i}}^{0}d\mathcal{N}_{1}\int_{\mathcal{N}_{i}}^{0}d\mathcal{N}_{2}\int_{\ln(k_{m})}^{\ln(k_{M})}\,d(\ln k)\frac{k}{H_{0}}H_{0}^{3}P(k)j_{l}[k\chi(\mathcal{N}_{1})]e^{-\mathcal{N}_{1}}e^{-\mathcal{N}_{2}}\nonumber \\
 &  & {}\times\{[\Sigma'(\mathcal{N}_{1})-\Sigma(\mathcal{N}_{1})]\,D(\mathcal{N}_{1})+\Sigma(\mathcal{N}_{1})D'(\mathcal{N}_{1})\}\phi^{\alpha}(\mathcal{N}_{2})D(\mathcal{N}_{2})b_{s}^{\alpha}j_{l}[k\chi(\mathcal{N}_{2})]\nonumber\\
 & \approx & -\left[6\pi^{2}\,\bar{\delta}_{H}^{2}\,\Omega_{m0}\,l_{12}^{-2}\right]\int_{\mathcal{N}_{i}}^{0}d\mathcal{N}e^{-\mathcal{N}}\,\frac{H}{H_{0}}\,\{[\Sigma'-\Sigma]\,D+\Sigma\,D'\}\phi^{\alpha}\,D\,b_{s}^{\alpha} {}\times[T_{m}(l_{12}H_{0}/\bar{\chi})]^{2}\left(\frac{l_{12}}{\bar{\chi}}\right)^{n_{s}}\,.
\end{eqnarray}
From Eq.\ (\ref{eq:CGG}), one can find a similar expression for $C_l^{\rm GG}$ by using the Limber approximation. This single integral can be numerically calculated, once more, by transforming it into an ODE, as follows:

\begin{eqnarray}
\frac{dD}{d\mathcal{N}} & = & \pi_{D}\,,\\
\frac{d\pi_{D}}{d\mathcal{N}} & = & -\left(2-\frac{3}{2}\,\Omega_{m}\right)\pi_{D}-\left[\frac{9}{2}\,\frac{Y\theta\,\Omega_{m}^{2}}{\left(Y\theta-2\right)^{2}}+\frac{3\Omega_{m}}{Y\theta-2}\right]D\,,\\
\frac{d\Omega_{m}}{d\mathcal{N}} & = & -3\Omega_{m}(1-\Omega_{m})\,,\\
\frac{dY}{d\mathcal{N}} & = & 3Y\Omega_{m}\,,\\
\frac{d\bar{\chi}}{d\mathcal{N}} & = & -\left(\frac{\Omega_{m}\sqrt{Y}}{\Omega_{m0}}\right)^{1/3}\,,\\
\frac{dZ_{G}^{\alpha}}{d\mathcal{N}} & = & \frac{e^{-\mathcal{N}}}{l_{12}^{2}\sqrt{Y}}\,[T_{m}(l_{12}H_{0}/\bar{\chi})]^{2}\left(\frac{l_{12}}{\bar{\chi}}\right)^{n_{s}+2}[\phi^{\alpha}]^{2}\,D^{2}\,,\\
\frac{dZ_{I}^{\alpha}}{d\mathcal{N}} & = & -\frac{e^{-\mathcal{N}}}{l_{12}^{2}\sqrt{Y}}\,[T_{m}(l_{12}H_{0}/\bar{\chi})]^{2}\left(\frac{l_{12}}{\bar{\chi}}\right)^{n_{s}}\{[\Sigma'-\Sigma]\,D+\Sigma\,\pi_{D}\}\phi^{\alpha}\,D\,,
\end{eqnarray}
where
\begin{equation}
\Sigma'=\frac{21}{2}\,\frac{\Omega_{m}\theta\,Y}{\left(Y\theta-2\right)^{3}}\left(Y\theta+\frac{12}{7}\,\Omega_{m}-2\right),
\end{equation}
and
\begin{eqnarray}
D(\mathcal{N}_{i}) & = & e^{\mathcal{N}_{i}}\,,\\
\pi_{D}(\mathcal{N}_{i}) & = & D'(\mathcal{N}_{i})=e^{\mathcal{N}_{i}}\,,\\
\Omega_{m}(\mathcal{N}_{i}) & = & \Omega_{mi}\,,\\
Y(\mathcal{N}_{i}) & = & Y_{i}\,,\\
\bar{\chi}(\mathcal{N}_{i}) & = & \bar{\chi}_{i}\,,\\
Z_{G}^{\alpha}(\mathcal{N}_{i}) & = & 0\,,\\
Z_{I}(\mathcal{N}_{i}) & = & 0\,,\\
C_{l}^{{\rm GG},\alpha L} & = & 2\pi^{2}\bar{\delta}_{H}^{2}\,(b_{s}^{\alpha})^{2}\,Z_{G}^{\alpha}\,,\\
C_{l}^{{\rm GI},\alpha L} & = & 6\pi^{2}\bar{\delta}_{H}^{2}b_{s}^{\alpha}\,\Omega_{m0}\,Z_{I}^{\alpha}\,.
\end{eqnarray}
%and we define $Y_{l}^{{\rm GG},\alpha L}=C_{l}^{{\rm GG},\alpha L}/(b_{s}^{\alpha})^{2}$, and $Y_{l}^{{\rm GI},\alpha L}=C_{l}^{{\rm GI},\alpha L}/b_{s}^{\alpha}$.

\begin{figure}
\includegraphics[width=9cm]{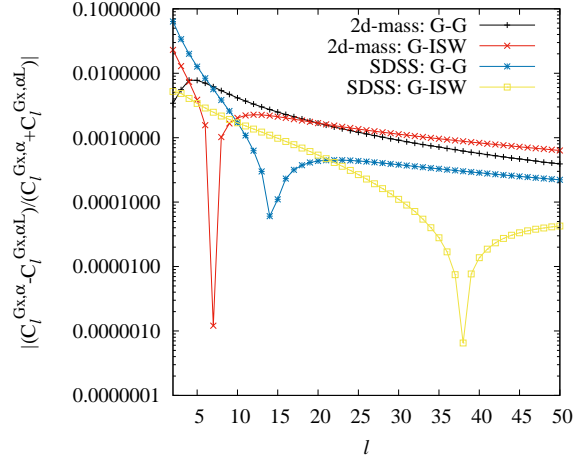}
\caption{\label{fig:Limber-approximation}Limber approximation at work for
large $l$. We plot here the relative error, that is $\frac{|C_{l}^{Gx,\alpha}-C_{l}^{Gx,\alpha L}|}{|C_{l}^{Gx,\alpha}+C_{l}^{Gx,\alpha L}|}$,
where $x\in\{G,I\}$, and $\alpha\in\{{\rm 2d{-}mass},{\rm SDSS}\}$.}
\end{figure}

We will use the Limber approximations in two different but useful ways. First of all, we use it to see whether for large $l$, the triple
integral gets closer and closer to the Limber results. Second, for $l>30$, since the approximation errors are, at worst, below 0.1\% we will use the Limber approximation result instead of the triple integral in order to save machine calculation time.  Figure \ref{fig:Limber-approximation} illustrates how well the Limber approximation works for large $l$.

\subsection{High $k$ dynamics for $\delta$}

For high $K$, the action for the perturbations (see e.g.\ \cite{DeFelice:2015moy,DeFelice:2016ufg})  in the normal branch reduces to
\begin{equation}
\mathcal{L}=\frac{1}{2}\,\Mpl^{2}\,N\,a^{3}\,\frac{3\Omega_{m}\,a^{2}}{K^{2}Y}\left[\left(\frac{\dot{\delta}}{N}\right)^{2}+\frac{3\left[4-\theta\,Y\left(3\Omega_{m}+2\right)\right]H^{2}\Omega_{m}\,\delta^{2}}{2\left(\theta Y-2\right)^{2}}\right],
\end{equation}
from which  the result of the self-accelerating branch and standard GR is recovered  in the limit $\theta\to0$. In this
case the dynamics for $\delta$ is given by
\begin{equation}
\delta''=-\left(2-\frac{3}{2}\,\Omega_{m}\right)\delta'-\left[\frac{9}{2}\,\frac{Y\theta\,\Omega_{m}^{2}}{\left(Y\theta-2\right)^{2}}+\frac{3\Omega_{m}}{Y\theta-2}\right]\delta\,.\label{eq:dynHighK}
\end{equation}
At early times, $Y\to0$ and $\Omega_{m}\to1$, so that
\begin{equation}
\delta''\approx-\frac{1}{2}\,\delta'+\frac{3}{2}\delta\,,
\end{equation}
which has the solution
\begin{equation}
\delta=C_{1}\,e^{\mathcal{N}}+C_{2}\,e^{-3\mathcal{N}/2}\approx C_{1}\,e^{\mathcal{N}}\,.
\end{equation}
Therefore at early times we have, as in GR, $\delta'\approx\delta$.

\section{Low $k$ regime}

In this section we evaluate the functions $\Sigma$ and $\Sigma'$, as well as the
differential equation for $\delta_{m}$ for all values of $k$ (i.e.\ not only the high-$k$
regime). Let us introduce 
\begin{equation}
K\equiv\frac{k}{H_{0}}\,,
\end{equation}
which is dimensionless.

\subsection{Action for the $\delta$ field}

We find that the Lagrangian for the $\delta$ field, for all values of $k$, in the normal branch, can be written
as (derived by using the results in \cite{DeFelice:2015moy}):
\begin{equation}
\mathcal{L}=\frac{1}{2}\,\Mpl^{2}\,N\,a^{3}\,Q\left[\left(\frac{\dot{\delta}}{N}\right)^{2}-\frac{9\left[K^{2}\theta\,\left(\Omega_{m}+\frac{2}{3}\right)Y+\frac{9}{2}\,\Omega_{m}^{2}a^{2}\theta-\frac{4}{3}\,K^{2}\right]H^{2}\Omega_{m}Y\,\delta^{2}}{2\left(Y\theta-2\right)^{2}\left(K^{2}Y+\frac{9}{2}\,\Omega_{m}a^{2}\right)}\right],\label{eq:azioDelta}
\end{equation}
where the no-ghost condition for the field $\delta$ can be written
as
\begin{equation}
Q=\frac{6\Omega_{m}\,a^{2}}{2\,K^{2}Y+9\,\Omega_{m}a^{2}}\,,
\end{equation}
which is always positive and well defined. Here $a=e^{N}=(\Omega_{m0}Y/\Omega_{m})^{1/3}$.
It should be noted that for values of $\bar{Y}$ for which $\bar{Y}=2/\theta$,
the mass of modes tends to blow up. In order to avoid such a possibility,
since $Y\leq 1$ (in the past, up to today), we require $\theta<2$.  The previous action Eq.~(\ref{eq:azioDelta}) reduces to the scalar-perturbation action of the self-accelerating branch and standard $\Lambda$CDM in GR in the limit $\theta\to0$.

\subsection{ODE for $\delta$}

Let us now study the differential equation for $\delta$. Here and in
the following a prime denotes a derivative with respect to
$\mathcal{N}$. We find that the ODE that determines the dynamics of
$\delta$ (the only independent scalar degree of freedom), can be
written as
\begin{equation}
\delta''+T_{2}\,\delta'+T_{3}\,\delta=0\,,\label{eq:eomD}
\end{equation}
where
\begin{eqnarray}
T_{2} & = & \frac{(8-6\,\Omega_{m})Y\,K^{2}-27\,\Omega_{m}a^{2}(\Omega_{m}-2)}{4\,K^{2}Y+18\,\Omega_{m}a^{2}}\,,\\
T_{3} & = & \frac{9Y\Omega_{m}\left\{ \left[\frac{1}{3}\,Y\,(3\,\Omega_{m}+2)\theta-\frac{4}{3}\right]K^{2}+\frac{9}{2}\,\Omega_{m}^{2}\,a^{2}\theta\right\} }{2\left(Y\theta-2\right)^{2}\left(K^{2}Y+\frac{9}{2}\,\Omega_{m}a^{2}\right)}\,.
\end{eqnarray}

\subsubsection{Initial conditions}

We will fix the initial condition for $\delta_{i}$ and $\delta'_{i}$
as follows. First of all let us introduce the gauge-invariant combination,
$\delta_{m}\equiv\delta\rho/\rho+3H\,v_{m}$, which can be written
in terms of the $\delta$ and $\delta'$ also as
\begin{equation}
\delta_{m}=\frac{2\left(K^{2}Y\theta+\frac{9}{2}\,\Omega_{m}\theta\,a^{2}-2\,K^{2}\right)Y\delta-6\,a^{2}\left(Y\theta-2\right)\delta'}{(2\,K^{2}Y+9\,\Omega_{m}a^{2})\,(Y\theta-2)}\,.
\end{equation}
Then we fix the amplitude of $\delta_{m}$ by setting the following
initial condition:
\begin{equation}
\delta_{mi}\equiv\delta_{m}(\mathcal{N}_{i})=a(\mathcal{N}_{i})=a_{i}=e^{\mathcal{N}_{i}}\,.\label{eq:ICdm}
\end{equation}
The other free initial condition will be set by demanding that the
theory, at high redshifts, reduces to GR in the matter domination era.
This requirement, in terms of $\psi'_{{\rm ISW}}$, is equivalent
to the initial condition: $\psi'_{{\rm ISW},i}=0$. This condition
ensures the integral of $C_{l}^{{\rm GI}}$ will vanish for $\mathcal{N}<\mathcal{N}_{i}$.
This gives the second required relation between the quantities $\delta_{i}$
and $\delta_{i}'$. Since we have, in general,
\begin{eqnarray}
\psi_{{\rm ISW}}' & = & Z_{{\rm ISW}}=\frac{9a^{2}\Omega_{m}}{2\left(K^{2}Y+\frac{9}{2}\,\Omega_{m}a^{2}\right)Y\left(Y\theta-2\right)^{3}K^{2}}\left\{ 24\,a^{2}\left(\Omega_{m}+\frac{2}{3}\right)\delta'\right.\nonumber \\
 &  & {}+\left[\frac{16}{3}\,K^{2}\left(\delta'-\delta\right)-9\,\left(6\,\delta\,\Omega_{m}^{3}+\Omega_{m}^{2}\left(\delta'-6\,\delta\right)+4\,\Omega_{m}\delta'+\frac{8}{3}\,\delta'\right)a^{2}\theta\right]Y\nonumber \\
 &  & {}-2\left[\left(6\,\Omega_{m}^{2}\,\delta+\Omega_{m}\left(\delta'-8\,\delta\right)+\frac{8}{3}(\delta'-\delta)\right)K^{2}-\frac{9}{4}\,\left(\Omega_{m}^{2}\left(\delta'-6\,\delta\right)+4\,\Omega_{m}\delta'+\frac{8}{3}\,\delta'\right)a^{2}\theta\right]\theta\,Y^{2}\nonumber \\
 &  & {}+\left.\left[\left(\Omega_{m}\left(\delta'-8\,\delta\right)+\frac{4}{3}\,(\delta'-\delta)\right)K^{2}-3\,a^{2}\left(\Omega_{m}+\frac{2}{3}\right)\theta\delta'\right]\theta^{2}Y^{3}\right\} ,\label{eq:evDpsiISW}
\end{eqnarray}
then we will impose
\begin{equation}
\psi'_{{\rm ISW}}(\mathcal{N}=\mathcal{N}_{i})=0\,.\label{eq:ICd_psi_ISW}
\end{equation}
Both initial conditions, namely Eq.\ (\ref{eq:ICdm}) and Eq.\ (\ref{eq:ICd_psi_ISW})
can be used to find the initial conditions for the variables $\delta_{i}$ and $\delta'_{i}$.

\section{The results}

We consider the observable to be the following quantity:
\begin{equation}
w^{\alpha}(\vartheta)=\frac{T_{{\rm CMB}}}{4\pi}\,\sum_{l}(2l+1)\,C_{l}^{{\rm GI},\alpha}P_{l}(\cos\vartheta)\,,
\end{equation}
where $T_{{\rm CMB}}=2.72548\times10^{6}$ $\mu$K and $\vartheta$ is the angle which denotes deviations from the center of the galaxy data set considered. The sum is evaluated in the range $2\leq l\leq150$, and the bias is found as described above.
Finally $C_{l}^{{\rm GI},\alpha}$ is found using either the triple integral (for $2\leq l\leq30$) or the Limber approximation. For both 2d-mass and SDSS, we have considered the data points from \cite{Giannantonio:2008zi} (using the jackknife errors estimation method).  Since the two data sets (from 2d-mass and SDSS) are uncorrelated (see e.g.\ Fig.\ 5 of \cite{Giannantonio:2008zi}), we can use their data points independently at the same time.

\begin{figure}[ht]
\includegraphics[height=8truecm]{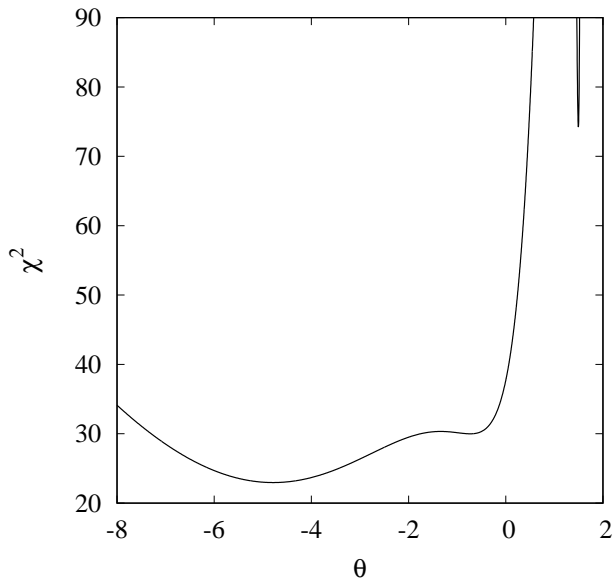}
\caption{\label{fig1}Total $\chi^2$ for different values of the normal-branch MTMG free parameter $\theta$.}
\end{figure}

In the following we show the results from the numerical triple integration/Limber approximation in order to find the above mentioned observable.  We fix the various parameters to the best fit value of Planck (in particular we set $\sigma_{8}(\mathcal{N}=-6)\approx 2.579\times10^{-3}$ for $\Lambda$CDM), but we let the parameter $\theta$ (proportional to the squared mass of the graviton) vary. The results are shown in Fig.\ \ref{fig1}, where we plot the total $\chi^2$ for the three experiments (2-dmass, SDSS and RSD) as a function of the normal-branch MTMG parameter $\theta$.
We find that at 1-$\sigma$, the $\chi^2$ has a minimum at $\theta\approx-4.79$ for which $\chi^2\approx22.97$. Considering a probability distribution of the kind $P\propto\exp(-\chi^2/2)$, we find that the 1-$\sigma$ contour is given by the following interval
\begin{equation}
\theta=-4.79^{{+}0.93}_{{-0.92}}\,.
\end{equation}
The value found for $\chi^2$ at $\theta=0$ (i.e.\ for $\Lambda$CDM) corresponds to $\chi^2(\theta=0)=37.68$. Although the $\chi^2$ for the normal branch of MTMG is lower than $\Lambda$CDM (or the self-accelerating branch of MTMG), the data still have quite large error bars. Furthermore we relied on the fact that we accepted the best-fit Planck values for the background parameters $h, \Omega_{m0}$. Nonetheless, we believe that future experiments, by increasing the sensitivity, will be able to distinguish between $\Lambda$CDM (or the self-accelerating branch of MTMG) and the normal branch of MTMG.  It should be noted that the graviton mass squared, $\mu^2$, is given by $\mu^2=\theta\,H_0^2$ (see e.g.\ \cite{DeFelice:2016ufg}). Hence, if the data confirm $\theta$ to be a negative number,  the tensor modes whose wavelength is of the order of the present horizon scale may grow now and in the future. However, even in such a case, when tensor mode instabilities arise, the time scale of the instability will be on the order of the age of the present universe. Nevertheless, it would then be worthwhile investigating how to probe such a growth with future experiments.

\begin{figure}[ht]
\includegraphics[height=8truecm]{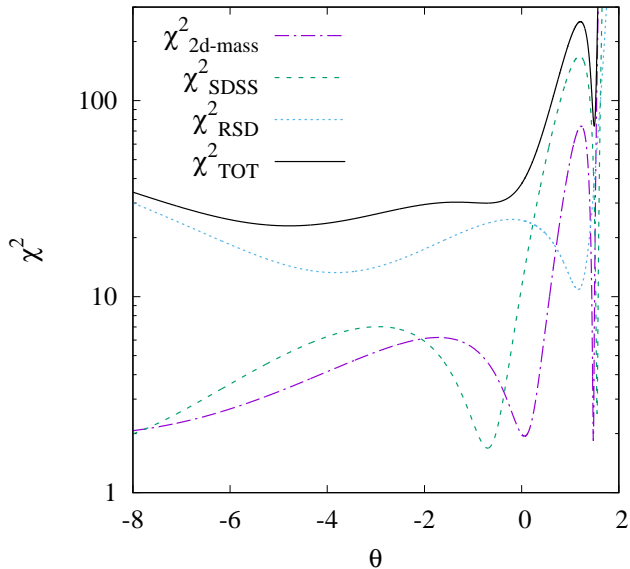}
\caption{\label{fig2}Contribution for the $\chi^2$ for each separate experiment. Here $\chi^2_{\rm TOT}$ corresponds to the $\chi^2$ shown in Fig.\ \ref{fig1}}
\end{figure}

\begin{figure}[ht]
\includegraphics[height=8truecm]{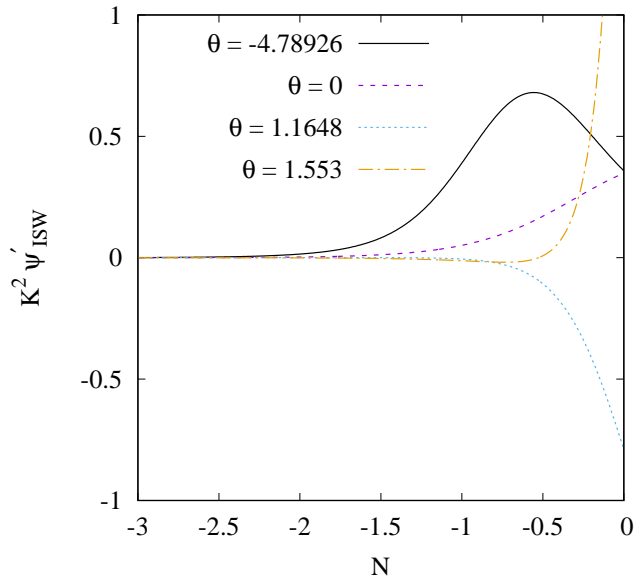}
\caption{\label{fig3}Comparison between the evolution of $K^2\psi_{\rm ISW}'$ for different values of $\theta$ in the high-$k$ regime.}
\end{figure}

Having a closer look at the results, one finds that for some positive values of $\theta$'s ($\theta\approx1.1$), anti-cross-correlations occur in the range where RSD data are well fit, so that the total $\chi^2$ reaches high values (see Fig.\ \ref{fig2}). This anti-cross-correlation effect stops for higher values of $\theta$ (SDSS data have a best-fit for $\theta\approx1.5$). For negative values of $\theta$, one finds that each separate contribution for the $\chi^2$ coming from each different experiment is consistently lower than the value found for $\Lambda$CDM with best-fit Planck parameters. This shows, given the chosen data sets, a tendency of the data to prefer a nonzero value of $\theta$ once we assume Planck best-fit values for the initial $\sigma_8$, etc. Each different contribution for the $\chi^2$ can be seen in Fig.\ \ref{fig2}.
As stated above, the theory for values of $\theta\simeq1.1$ tends to have a bad fit to the ISW-galaxy cross-correlation data. This happens, as shown in Fig.~\ref{fig3}, because the quantity $\psi_{\rm ISW}'$ acquires an opposite sign (compared to $\Lambda$CDM) at low redshifts. This phenomenon gives rise to the anti-cross-correlation. However, for other values of $\theta$, positive cross-correlation is achieved, like in GR.

\begin{figure}[ht]
\includegraphics[height=7truecm]{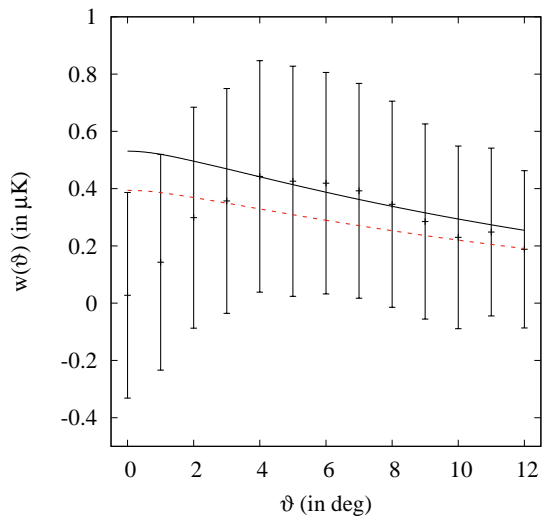}
\caption{\label{fig4}Comparison between the best-fit model of the normal branch of MTMG (black solid line) vs. $\Lambda$CDM (or the self-accelerating branch of MTMG) (red dashed line) fitting the 2dmass data.}
\end{figure}

\begin{figure}[ht]
\includegraphics[height=8truecm]{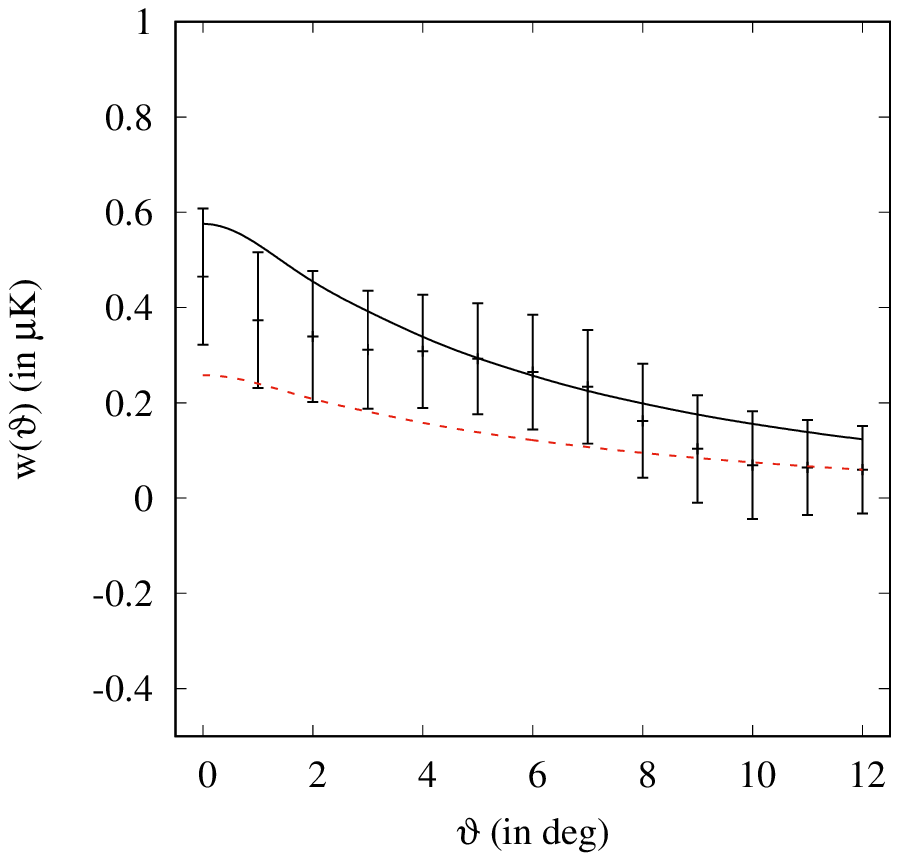}
\caption{\label{fig5}Comparison between the best-fit model of the normal branch of MTMG (black solid line) vs. $\Lambda$CDM (or the self-accelerating branch of MTMG) (red dashed line) fitting the SDSS data.}
\end{figure}

In Fig.~ \ref{fig4} and \ref{fig5}, we show how the best-fit model in the normal branch of MTMG fits the data compared to $\Lambda$CDM (or the self-accelerating branch of MTMG).

\section{Conclusions}

In this work we studied the influence of the normal-branch MTMG growth evolution on the ISW-galaxy cross-correlation observable. Then we performed a fit to existing data to check whether anti-cross-correlation would rule out the allowed RSD data parameter space for the normal branch of MTMG studied in \cite{DeFelice:2016ufg}. We performed a full triple integral to calculate the ISW-galaxy cross-correlations, so that we could take into account the deviation from GR (or the self-accelerating branch of MTMG) not only at small scales but also at large scales, at which ISW-data are quite sensitive. We also checked that the Limber approximation is very precise for values of $l\geq30$ (to 0.1\% at worst).

We chose ISW-galaxy correlation data for which the least amount of priors were needed to define the bias for the experiments. In particular the data selected allowed us to neglect any possible time dependence of the bias thanks to the window function for the galaxy data being peaked around a particular value of the redshift. We succeeded to reproduce the $\Lambda$CDM fit to the existing data, and then we expanded the study of the fit to nonzero values of $\theta$, the parameter proportional to $\mu^2/H_0^2$; $\mu$ being the mass of the tensor modes for the normal branch of MTMG. The predictions of the theory reduce to that of $\Lambda$CDM (or the self-accelerating branch of MTMG) for $\theta=0$.

We find that the ISW-galaxy cross-correlation data disfavors a region of $\theta \simeq 1.1$, where anti-cross-correlation occurs. This happens as $\psi_{\rm ISW}'$ changes sign (with respect to the $\Lambda$CDM behavior) at low redshifts. However, for other values of $\theta$ positive correlation does take place, for example for $\theta\leq0$ and $\theta>1.5$.  If we further fix both $h$ and $\Omega_{m0}$ to the best-fit Planck values, the $\chi^2$ analysis for ISW-galaxy data combined with the chosen RSD data suggests that negative values of $\theta$ give a better fit. However, this final result is probably not enough to state anything definitive. Especially, as is clear from Fig.~\ref{fig2}, the normal branch of MTMG (as well as $\Lambda$CDM in GR and the self-accelerating branch of MTMG) does not solve the tensions completely. Nonetheless, if future experiments point more and more toward the need of weaker gravity to explain the RSD data, then indeed, the normal branch of MTMG could be distinguished from $\Lambda$CDM in GR (or the self-accelerating branch of MTMG).

\begin{acknowledgments}
We thank Tommaso Giannantonio, Atsushi Taruya and Takahiro Tanaka for useful comments. ADF was supported by JSPS KAKENHI Grant Numbers 16K05348, 16H01099.  The work of SM was supported by Japan Society for the Promotion of Science (JSPS) Grants-in-Aid for Scientific Research (KAKENHI) No.\ 17H02890, No.\ 17H06359, and by World Premier International Research Center Initiative (WPI), MEXT, Japan. NB acknowledges funding from the European Research Council under the European Union's Seventh Framework Programme (FP7/2007-2013)/ERC Grant Agreement No. 617656 ``Theories and Models of the Dark Sector: Dark Matter, Dark Energy and Gravity.'' NB also  acknowledges support from the Central European Institute for Cosmology and Fundamental Physics (CEICO) in Prague, where part of this work was conducted.
\end{acknowledgments}


\begin{thebibliography}{10}
\bibitem{Hinshaw:2012aka} 
  G.~Hinshaw {\it et al.} [WMAP Collaboration],
  %``Nine-Year Wilkinson Microwave Anisotropy Probe (WMAP) Observations: Cosmological Parameter Results,''
  Astrophys.\ J.\ Suppl.\  {\bf 208}, 19 (2013)
  doi:10.1088/0067-0049/208/2/19
  [arXiv:1212.5226 [astro-ph.CO]].
  
\bibitem{Bennett:2012zja} 
  C.~L.~Bennett {\it et al.} [WMAP Collaboration],
  %``Nine-Year Wilkinson Microwave Anisotropy Probe (WMAP) Observations: Final Maps and Results,''
  Astrophys.\ J.\ Suppl.\  {\bf 208}, 20 (2013)
  doi:10.1088/0067-0049/208/2/20
  [arXiv:1212.5225 [astro-ph.CO]].
  
\bibitem{Ade:2015xua} 
  P.~A.~R.~Ade {\it et al.} [Planck Collaboration],
  %``Planck 2015 results. XIII. Cosmological parameters,''
  Astron.\ Astrophys.\  {\bf 594}, A13 (2016)
  doi:10.1051/0004-6361/201525830
  [arXiv:1502.01589 [astro-ph.CO]].

\bibitem{Aubourg:2014yra} 
  \'E.~Aubourg {\it et al.},
  %``Cosmological implications of baryon acoustic oscillation measurements,''
  Phys.\ Rev.\ D {\bf 92}, no. 12, 123516 (2015)
  doi:10.1103/PhysRevD.92.123516
  [arXiv:1411.1074 [astro-ph.CO]].

%\bibitem{Wang:2016wjr}
%Yuting Wang et~al.
%\newblock {The clustering of galaxies in the completed SDSS-III Baryon
%  Oscillation Spectroscopic Survey: tomographic BAO analysis of DR12 combined
%  sample in configuration space}.
%\newblock {\em Mon. Not. Roy. Astron. Soc.}, 469(3):3762--3774, 2017.

\bibitem{Cuesta:2015mqa} 
  A.~J.~Cuesta {\it et al.},
  %``The clustering of galaxies in the SDSS-III Baryon Oscillation Spectroscopic Survey: Baryon Acoustic Oscillations in the correlation function of LOWZ and CMASS galaxies in Data Release 12,''
  Mon.\ Not.\ Roy.\ Astron.\ Soc.\  {\bf 457}, no. 2, 1770 (2016)
  doi:10.1093/mnras/stw066
  [arXiv:1509.06371 [astro-ph.CO]].

\bibitem{Wang:2016wjr} 
  Y.~Wang {\it et al.} [BOSS Collaboration],
  %``The clustering of galaxies in the completed SDSS-III Baryon Oscillation Spectroscopic Survey: tomographic BAO analysis of DR12 combined sample in configuration space,''
  Mon.\ Not.\ Roy.\ Astron.\ Soc.\  {\bf 469}, no. 3, 3762 (2017)
  doi:10.1093/mnras/stx1090
  [arXiv:1607.03154 [astro-ph.CO]].

%\bibitem{Cuesta:2015mqa}
%Antonio~J. Cuesta et~al.
%\newblock {The clustering of galaxies in the SDSS-III Baryon Oscillation
%  Spectroscopic Survey: Baryon Acoustic Oscillations in the correlation
%  function of LOWZ and CMASS galaxies in Data Release 12}.
%\newblock {\em Mon. Not. Roy. Astron. Soc.}, 457(2):1770--1785, 2016.

\bibitem{Paris:2017xme} 
  I.~P\^aris {\it et al.} [SDSS Collaboration],
  %``The Sloan Digital Sky Survey Quasar Catalog: Fourteenth Data Release,''
  arXiv:1712.05029 [astro-ph.GA].
  
\bibitem{Blake:2011rj} 
  C.~Blake {\it et al.},
  %``The WiggleZ Dark Energy Survey: the growth rate of cosmic structure since redshift z=0.9,''
  Mon.\ Not.\ Roy.\ Astron.\ Soc.\  {\bf 415}, 2876 (2011)
  doi:10.1111/j.1365-2966.2011.18903.x
  [arXiv:1104.2948 [astro-ph.CO]].

\bibitem{Bernal:2016gxb} 
  J.~L.~Bernal, L.~Verde and A.~G.~Riess,
  %``The trouble with $H_0$,''
  JCAP {\bf 1610}, no. 10, 019 (2016)
  doi:10.1088/1475-7516/2016/10/019
  [arXiv:1607.05617 [astro-ph.CO]].
  
\bibitem{Cuesta:2014asa} 
  A.~J.~Cuesta, L.~Verde, A.~Riess and R.~Jimenez,
  %``Calibrating the cosmic distance scale ladder: the role of the sound horizon scale and the local expansion rate as distance anchors,''
  Mon.\ Not.\ Roy.\ Astron.\ Soc.\  {\bf 448}, no. 4, 3463 (2015)
  doi:10.1093/mnras/stv261
  [arXiv:1411.1094 [astro-ph.CO]].
  
%\bibitem{Bernal:2016gxb}
%Jose~Luis Bernal, Licia Verde, and Adam~G. Riess.
%\newblock {The trouble with $H_0$}.
%\newblock {\em JCAP}, 1610(10):019, 2016.

\bibitem{DeFelice:2016ufg} 
  A.~De Felice and S.~Mukohyama,
  %``Graviton mass might reduce tension between early and late time cosmological data,''
  Phys.\ Rev.\ Lett.\  {\bf 118}, no. 9, 091104 (2017)
  doi:10.1103/PhysRevLett.118.091104
  [arXiv:1607.03368 [astro-ph.CO]].
%\bibitem{DeFelice:2016ufg}
%Antonio De~Felice and Shinji Mukohyama.
%\newblock {Graviton mass might reduce tension between early and late time cosmological data}.
%\newblock {\em Phys. Rev. Lett.}, 118(9):091104, 2017.

\bibitem{DeFelice:2015hla} 
  A.~De Felice and S.~Mukohyama,
  %``Minimal theory of massive gravity,''
  Phys.\ Lett.\ B {\bf 752}, 302 (2016)
  doi:10.1016/j.physletb.2015.11.050
  [arXiv:1506.01594 [hep-th]].
%\bibitem{DeFelice:2015hla}
%Antonio De~Felice and Shinji Mukohyama.
%\newblock {Minimal theory of massive gravity}.
%\newblock {\em Phys. Lett.}, B752:302--305, 2016.

\bibitem{DeFelice:2015moy} 
  A.~De Felice and S.~Mukohyama,
  %``Phenomenology in minimal theory of massive gravity,''
  JCAP {\bf 1604}, no. 04, 028 (2016)
  doi:10.1088/1475-7516/2016/04/028
  [arXiv:1512.04008 [hep-th]].
%\bibitem{DeFelice:2015moy}
%Antonio De~Felice and Shinji Mukohyama.
%\newblock {Phenomenology in minimal theory of massive gravity}.
%\newblock {\em JCAP}, 1604(04):028, 2016.

\bibitem{deRham:2010ik} 
  C.~de Rham and G.~Gabadadze,
  %``Generalization of the Fierz-Pauli Action,''
  Phys.\ Rev.\ D {\bf 82}, 044020 (2010)
  doi:10.1103/PhysRevD.82.044020
  [arXiv:1007.0443 [hep-th]].  
%\bibitem{deRham:2010ik}
%Claudia de~Rham and Gregory Gabadadze.
%\newblock {Generalization of the Fierz-Pauli Action}.
%\newblock {\em Phys. Rev.}, D82:044020, 2010.

%\bibitem{deRham:2010kj}
%Claudia de~Rham, Gregory Gabadadze, and Andrew~J. Tolley.
%\newblock {Resummation of Massive Gravity}.
%\newblock {\em Phys. Rev. Lett.}, 106:231101, 2011.

\bibitem{deRham:2010kj} 
  C.~de Rham, G.~Gabadadze and A.~J.~Tolley,
  %``Resummation of Massive Gravity,''
  Phys.\ Rev.\ Lett.\  {\bf 106}, 231101 (2011)
  doi:10.1103/PhysRevLett.106.231101
  [arXiv:1011.1232 [hep-th]].

\bibitem{DeFelice:2012mx} 
  A.~De Felice, A.~E.~Gumrukcuoglu and S.~Mukohyama,
  %``Massive gravity: nonlinear instability of the homogeneous and isotropic universe,''
  Phys.\ Rev.\ Lett.\  {\bf 109}, 171101 (2012)
  doi:10.1103/PhysRevLett.109.171101
  [arXiv:1206.2080 [hep-th]].
%\bibitem{DeFelice:2012mx}
%Antonio De~Felice, A.~Emir Gumrukcuoglu, and Shinji Mukohyama.
%\newblock {Massive gravity: nonlinear instability of the homogeneous and isotropic universe}.
%\newblock {\em Phys. Rev. Lett.}, 109:171101, 2012.

\bibitem{DAmico:2011eto} 
  G.~D'Amico, C.~de Rham, S.~Dubovsky, G.~Gabadadze, D.~Pirtskhalava and A.~J.~Tolley,
  %``Massive Cosmologies,''
  Phys.\ Rev.\ D {\bf 84}, 124046 (2011)
  doi:10.1103/PhysRevD.84.124046
  [arXiv:1108.5231 [hep-th]].

\bibitem{Gumrukcuoglu:2012aa} 
  A.~E.~Gumrukcuoglu, C.~Lin and S.~Mukohyama,
  %``Anisotropic Friedmann-Robertson-Walker universe from nonlinear massive gravity,''
  Phys.\ Lett.\ B {\bf 717}, 295 (2012)
  doi:10.1016/j.physletb.2012.09.049
  [arXiv:1206.2723 [hep-th]].
  
\bibitem{DAmico:2012hia} 
  G.~D'Amico, G.~Gabadadze, L.~Hui and D.~Pirtskhalava,
  %``Quasidilaton: Theory and cosmology,''
  Phys.\ Rev.\ D {\bf 87}, 064037 (2013)
  doi:10.1103/PhysRevD.87.064037
  [arXiv:1206.4253 [hep-th]].

\bibitem{Huang:2012pe} 
  Q.~G.~Huang, Y.~S.~Piao and S.~Y.~Zhou,
  %``Mass-Varying Massive Gravity,''
  Phys.\ Rev.\ D {\bf 86}, 124014 (2012)
  doi:10.1103/PhysRevD.86.124014
  [arXiv:1206.5678 [hep-th]].

%\cite{Cheung:2016yqr}
\bibitem{Cheung:2016yqr} 
  C.~Cheung and G.~N.~Remmen,
  %``Positive Signs in Massive Gravity,''
  JHEP {\bf 1604}, 002 (2016)
  doi:10.1007/JHEP04(2016)002
  [arXiv:1601.04068 [hep-th]].

%\cite{Bonifacio:2016wcb}
\bibitem{Bonifacio:2016wcb} 
  J.~Bonifacio, K.~Hinterbichler and R.~A.~Rosen,
  %``Positivity constraints for pseudolinear massive spin-2 and vector Galileons,''
  Phys.\ Rev.\ D {\bf 94}, no. 10, 104001 (2016)
  doi:10.1103/PhysRevD.94.104001
  [arXiv:1607.06084 [hep-th]].
  
%\cite{Bellazzini:2017fep}
\bibitem{Bellazzini:2017fep} 
  B.~Bellazzini, F.~Riva, J.~Serra and F.~Sgarlata,
  %``Beyond Positivity Bounds and the Fate of Massive Gravity,''
  Phys.\ Rev.\ Lett.\  {\bf 120}, no. 16, 161101 (2018)
  doi:10.1103/PhysRevLett.120.161101
  [arXiv:1710.02539 [hep-th]].  

%\cite{deRham:2017xox}
\bibitem{deRham:2017xox} 
  C.~de Rham, S.~Melville and A.~J.~Tolley,
  %``Improved Positivity Bounds and Massive Gravity,''
  JHEP {\bf 1804}, 083 (2018)
  doi:10.1007/JHEP04(2018)083
  [arXiv:1710.09611 [hep-th]].

\bibitem{Afshordi:2003xu} 
  N.~Afshordi, Y.~S.~Loh and M.~A.~Strauss,
  %``Cross - correlation of the Cosmic Microwave Background with the 2MASS galaxy survey: Signatures of dark energy, hot gas, and point sources,''
  Phys.\ Rev.\ D {\bf 69}, 083524 (2004)
  doi:10.1103/PhysRevD.69.083524
  [astro-ph/0308260].

\bibitem{Afshordi:2004kz} 
  N.~Afshordi,
  %``Integrated Sachs-Wolfe effect in cross - correlation: The Observer's manual,''
  Phys.\ Rev.\ D {\bf 70}, 083536 (2004)
  doi:10.1103/PhysRevD.70.083536
  [astro-ph/0401166].
  
\bibitem{Corasaniti:2005pq} 
  P.~S.~Corasaniti, T.~Giannantonio and A.~Melchiorri,
  %``Constraining dark energy with cross-correlated CMB and large scale structure data,''
  Phys.\ Rev.\ D {\bf 71}, 123521 (2005)
  doi:10.1103/PhysRevD.71.123521
  [astro-ph/0504115].
  
\bibitem{Pogosian:2005ez} 
  L.~Pogosian, P.~S.~Corasaniti, C.~Stephan-Otto, R.~Crittenden and R.~Nichol,
  %``Tracking dark energy with the ISW effect: Short and long-term predictions,''
  Phys.\ Rev.\ D {\bf 72}, 103519 (2005)
  doi:10.1103/PhysRevD.72.103519
  [astro-ph/0506396].
  
\bibitem{Giannantonio:2008zi} 
  T.~Giannantonio, R.~Scranton, R.~G.~Crittenden, R.~C.~Nichol, S.~P.~Boughn, A.~D.~Myers and G.~T.~Richards,
  %``Combined analysis of the integrated Sachs-Wolfe effect and cosmological implications,''
  Phys.\ Rev.\ D {\bf 77}, 123520 (2008)
  doi:10.1103/PhysRevD.77.123520
  [arXiv:0801.4380 [astro-ph]].

\bibitem{Kimura:2011td} 
  R.~Kimura, T.~Kobayashi and K.~Yamamoto,
  %``Observational Constraints on Kinetic Gravity Braiding from the Integrated Sachs-Wolfe Effect,''
  Phys.\ Rev.\ D {\bf 85}, 123503 (2012)
  doi:10.1103/PhysRevD.85.123503
  [arXiv:1110.3598 [astro-ph.CO]].

\bibitem{Tommaso:2012}
T.~Giannantonio, R.~Crittenden, R.~Nichol, and A.~J.~Ross,
%"The significance of the integrated Sachs–Wolfe effect revisited,"
  Mon.\ Not.\ Roy.\ Astron.\ Soc.\  {\bf 426}, 2581 (2012)
  doi:10.1111/j.1365-2966.2012.21896.x
  [arXiv:1209.2125  [astro-ph]]
  
\bibitem{Giannantonio:2013kqa} 
  T.~Giannantonio and W.~J.~Percival,
  %``Using correlations between CMB lensing and large-scale structure to measure primordial non-Gaussianity,''
  Mon.\ Not.\ Roy.\ Astron.\ Soc.\  {\bf 441}, L16 (2014)
  doi:10.1093/mnrasl/slu036
  [arXiv:1312.5154 [astro-ph.CO]].


%\bibitem{Giannantonio:2008zi}
%Tommaso Giannantonio, Ryan Scranton, Robert~G. Crittenden, Robert~C. Nichol,
%  Stephen~P. Boughn, Adam~D. Myers, and Gordon~T. Richards.
%\newblock {Combined analysis of the integrated Sachs-Wolfe effect and cosmological implications}.
%\newblock {\em Phys. Rev.}, D77:123520, 2008.

%\bibitem{Corasaniti:2005pq}
%Pier-Stefano Corasaniti, Tommaso Giannantonio, and Alessandro Melchiorri.
%\newblock {Constraining dark energy with cross-correlated CMB and large scale structure data}.
%\newblock {\em Phys. Rev.}, D71:123521, 2005.

%\bibitem{Boulware:1973my}
%D.~G. Boulware and Stanley Deser.
%\newblock {Can gravitation have a finite range?}
%\newblock {\em Phys. Rev.}, D6:3368--3382, 1972.

%\bibitem{DAmico:2011eto}
%G.~D'Amico, C.~de~Rham, S.~Dubovsky, G.~Gabadadze, D.~Pirtskhalava, and A.~J.
%  Tolley.
%\newblock {Massive Cosmologies}.
%\newblock {\em Phys. Rev.}, D84:124046, 2011.

%\bibitem{Ade:2015xua}
%P.~A.~R. Ade et~al.
%\newblock {Planck 2015 results. XIII. Cosmological parameters}.
%\newblock {\em Astron. Astrophys.}, 594:A13, 2016.

%\bibitem{Hinshaw:2012aka}
%G.~Hinshaw et~al.
%\newblock {Nine-Year Wilkinson Microwave Anisotropy Probe (WMAP) Observations:
%  Cosmological Parameter Results}.
%\newblock {\em Astrophys. J. Suppl.}, 208:19, 2013.

%\bibitem{Bennett:2012zja}
%C.~L. Bennett et~al.
%\newblock {Nine-Year Wilkinson Microwave Anisotropy Probe (WMAP) Observations:
%  Final Maps and Results}.
%\newblock {\em Astrophys. J. Suppl.}, 208:20, 2013.

%\bibitem{Cuesta:2014asa}
%Antonio~J. Cuesta, Licia Verde, Adam Riess, and Raul Jimenez.
%\newblock {Calibrating the cosmic distance scale ladder: the role of the sound
%  horizon scale and the local expansion rate as distance anchors}.
%\newblock {\em Mon. Not. Roy. Astron. Soc.}, 448(4):3463--3471, 2015.

%\bibitem{Aubourg:2014yra}
%Éric Aubourg et~al.
%\newblock {Cosmological implications of baryon acoustic oscillation
%  measurements}.
%\newblock {\em Phys. Rev.}, D92(12):123516, 2015.

\bibitem{Eisenstein:1997ik} 
  D.~J.~Eisenstein and W.~Hu,
  %``Baryonic features in the matter transfer function,''
  Astrophys.\ J.\  {\bf 496}, 605 (1998)
  doi:10.1086/305424
  [astro-ph/9709112].
  
\bibitem{Eisenstein:1997jh} 
  D.~J.~Eisenstein and W.~Hu,
  %``Power spectra for cold dark matter and its variants,''
  Astrophys.\ J.\  {\bf 511}, 5 (1999)
  doi:10.1086/306640
  [astro-ph/9710252].

\bibitem{Prat:2016xor} 
  J.~Prat {\it et al.} [DES Collaboration],
  %``Galaxy bias from galaxy–galaxy lensing in the DES science verification data,''
  Mon.\ Not.\ Roy.\ Astron.\ Soc.\  {\bf 473}, no. 2, 1667 (2018)
  doi:10.1093/mnras/stx2430
  [arXiv:1609.08167 [astro-ph.CO]].
  %%CITATION = doi:10.1093/mnras/stx2430;%%

\bibitem{Biasrep}
V.~Desjacques, D.~Jeong, F.~Schmidt,
%"Large-scale galaxy bias,"
Physics Reports, 2018, ISSN 0370-1573,
doi:10.1016/j.physrep.2017.12.002.




\end{thebibliography}
\end{document}